# Galaxy build-up in the first 1.5 Gyr of cosmic history: insights from the stellar mass function at z ∼ 4–9 from *JWST* NIRCam observations

Andrea Weibel,[1]★ Pascal A. Oesch,[1,2,3] Laia Barrufet,[1,4] Rashmi Gottumukkala,[1,2,3]
Richard S. Ellis,[5] Paola Santini,[6] John R. Weaver,[7] Natalie Allen,[2,3] Rychard Bouwens,[8]
Rebecca A. A. Bowler,[9] Gabe Brammer,[2,3] Adam C. Carnall,[4] Fergus Cullen,[4] Pratika Dayal,[10]
Mark Dickinson,[11] Callum T. Donnan,[4] James S. Dunlop,[4] Mauro Giavalisco,[7] Norman A. Grogin,[12]
Garth D. Illingworth,[13] Anton M. Koekemoer,[12] Ivo Labbe,[14] Danilo Marchesini,[15]
Derek J. McLeod,[4] Ross J. McLure,[4] Rohan P. Naidu,[16]† Pablo G. Pérez-González,[17] Marko Shuntov,[2,3]
Mauro Stefanon,[18,19] Sune Toft[2,3] and Mengyuan Xiao[1]

*Affiliations are listed at the end of the paper*

## ABSTRACT

Combining the public *JWST*/NIRCam imaging programs CEERS, PRIMER, and JADES, spanning a total area of ∼ 500 arcmin$^2$, we obtain a sample of >30 000 galaxies at $z_{\rm phot}$ ∼ 4–9 that allows us to perform a complete, rest-optical-selected census of the galaxy population at $z > 3$. Comparing the stellar mass $M_*$ and the UV-slope $\beta$ distributions between *JWST*- and *HST*-selected samples, we generally find very good agreement and no significant biases. Nevertheless, *JWST* enables us to probe a new population of UV-red galaxies that was missing from previous *HST*-based Lyman-break galaxy (LBG) samples. We measure galaxy stellar mass functions (SMFs) at $z$ ∼ 4–9 down to limiting masses of $10^{7.5}$–$10^{8.5}\,{\rm M}_\odot$, finding steep low-mass slopes over the entire redshift range, reaching values of $\alpha \approx -2$ at $z \gtrsim 6$. At the high-mass end, UV-red galaxies dominate at least out to $z$ ∼ 6. The implied redshift evolution of the SMF suggests a rapid build-up of massive dust-obscured or quiescent galaxies from $z$ ∼ 6 to $z$ ∼ 4 as well as an enhanced efficiency of star formation towards earlier times ($z \gtrsim 6$). Finally, we show that the galaxy mass density grows by a factor ∼ 20× from $z$ ∼ 9 to $z$ ∼ 4. Our results emphasize the importance of rest-frame optically selected samples in inferring accurate distributions of physical properties and studying the mass build-up of galaxies in the first 1.5 Gyr of cosmic history.

## 1 INTRODUCTION

For roughly three decades now, astronomers have used the Lyman-break technique to identify the so-called Lyman-break galaxies (LBGs) at $z \gtrsim 3$ (e.g. Madau et al. 1996; Steidel et al. 1996, see Giavalisco 2002; Shapley 2011 for reviews), initially from ground-based broad-band photometry, and later in particular with the *Hubble Space Telescope* (*HST*). The Advanced Camera for Surveys (ACS) which probes the optical part of the electromagnetic spectrum at $\lambda \approx 0.3$–0.9 μm enabled efficient selection of LBGs at $z$ ∼ 3–6 (e.g. Bunker et al. 2004; Dickinson et al. 2004; Giavalisco et al. 2004b; Beckwith et al. 2006; Coe et al. 2006; Bouwens et al. 2007). In 2009, the Wide Field Camera 3 (WFC3) was installed on the *HST*, extending its accessible wavelength range into the near-infrared, out to ≈ 1.6 μm and allowing for the discovery of LBGs out to $z$ ∼ 11 (e.g. Finkelstein et al. 2010; Bouwens et al. 2010b; Hathi et al. 2010; Lorenzoni et al. 2011; McLure et al. 2011; Ellis et al. 2013; Oesch et al. 2016). While studies at those extreme redshifts remained limited to small numbers of objects (e.g. Bouwens et al. 2016; McLeod, McLure & Dunlop 2016; Ishigaki et al. 2018; Oesch et al. 2018), much larger samples containing thousands of galaxies have been compiled at redshifts $z$ ∼ 4–8, both with *HST* and ground-based imaging, allowing for robust constraints on the UV-luminosity function at those epochs (e.g. Bouwens et al. 2015, 2021; Finkelstein et al. 2015; Ono et al. 2018). Unfortunately, the rest-frame optical part of the spectrum shifts out of the reddest WFC3 filter, the *H* band or *F*160*W*, at a redshift of $z$ ∼ 3, meaning that the selection of *HST*-based LBGs, and their inferred properties are solely based on the rest-frame UV part of their spectrum.

Inferring the stellar mass of a galaxy from photometry strongly depends on the dust extinction, the stellar initial mass function (IMF), the metallicity, and the assumed star formation history (SFH; see Conroy 2013 for a review). Having access to the rest-frame optical emission that traces the light of the most common low-mass stars significantly improves measurements of the stellar mass as it helps to put tighter constraints on all the mentioned parameters (e.g. Stefanon et al. 2017). Various authors have therefore complemented *HST*-selected LBG samples with data from the *Spitzer*/IRAC instrument, probing wavelengths of 3–10 μm, and/or ground-based $K_S$-band imaging to constrain the stellar mass function (SMF) at $z \gtrsim 3$ (e.g. Grazian et al. 2015; Song et al. 2016; Stefanon et al. 2017, 2021; Bhatawdekar et al. 2019; Kikuchihara et al. 2020; McLeod et al. 2021; Weaver et al. 2023), which is a fundamental measurement of the evolution of the galaxy population and provides an important observational benchmark to compare with simulations.

★ E-mail: andrea.weibel@unige.ch
† NASA Hubble Fellow

In parallel, astronomers have been discovering and investigating galaxies that were missing completely from LBG samples due to their faintness at observed optical to NIR wavelengths, causing them to remain undetected even in the deepest *HST*- and ground-based surveys. Initially, galaxies that had no counterpart in the optical or NIR were detected at sub-mm wavelengths with the Submillimetre Common User Bolometer Array (Holland et al. 1999; see e.g. Dunlop et al. 2004; Chapman et al. 2005). Similar sources were identified with the SPIRE instrument onboard the *Herschel* telescope (e.g. Casey et al. 2012). At later times, the Atacama Large Millimeter/submillimeter Array (ALMA) enabled the detection of additional extremely red sources that were lacking optical or NIR counterparts (e.g. Simpson et al. 2014; Williams et al. 2019; Yamaguchi et al. 2019; Fudamoto et al. 2021). A complementary approach was to select galaxies showing a very red colour between *Spitzer*/IRAC imaging at 3–4 μm and the reddest *HST*/WFC3 filter, the *H* band at 1.6 μm (e.g. Huang et al. 2011; Wang et al. 2016, 2019; Alcalde Pampliega et al. 2019; Xiao et al. 2023).

While such extremely red galaxies were rare, they were argued to be very massive, highly star-forming, and dust-obscured systems, contributing significantly to the star formation rate (SFR) density of the Universe at $z \gtrsim 3$ (e.g. Zavala et al. 2021). They were also shown to contribute significantly to the high-mass end of the SMF at $z \gtrsim 3$ (e.g. Caputi et al. 2011, 2015; Stefanon et al. 2015). Studies of these galaxies were however limited by small survey areas, low sensitivity, and/or poor spatial resolution at wavelengths beyond the reach of *HST*.

Recently, the Near Infrared Camera (NIRCam, Rieke, Kelly & Horner 2005) onboard the *JWST* has extended our view of the cosmos to wavelengths of up to 5 μm at a sensitivity and spatial resolution comparable to or exceeding that of *HST*. It therefore opens a new window on the rest-optical emission of galaxies at redshifts out to $z \sim 9$. Selecting galaxies from NIRCam imaging via a red colour between $\lambda_{obs} \sim 1.5$ μm and $\sim 4.5$ μm, various studies have confirmed the dusty star-forming nature of the high-redshift ($z \gtrsim 3$) galaxies in the resulting samples, as well as their significant contribution to the cosmic stellar mass and SFR density (e.g. Barrufet et al. 2023; Gómez-Guijarro et al. 2023; Nelson et al. 2023; Pérez-González et al. 2023; Rodighiero et al. 2023; Gottumukkala et al. 2024; Williams et al. 2024).

On the other hand, LBG samples are by definition missing such sources, and are generally biased towards UV-bright sources and blue UV-slopes $\beta$ (where $f_\lambda \propto \lambda^\beta$). For example, Bouwens et al. (2012) showed that LBGs at $z \sim 4$–7 were generally very blue with a mean $\beta \sim -2$, and a trend of bluer slopes towards lower UV-luminosities as well as towards higher redshifts (see also Bouwens et al. 2010a; Finkelstein et al. 2010, 2012; McLure et al. 2011; Dunlop et al. 2012). The question remained, how many 'red' galaxies with UV-slopes of e.g. $\beta \gtrsim -1.2$ there might be, and what is their contribution to the galaxy census at $z \gtrsim 3$. With the newly available NIRcam imaging, we can now attempt to answer this question.

In this work, we focus on the redshift range $z \sim 4$–9 spanning $\sim 1$ Gyr of cosmic time, from $\sim 0.5$ to $\sim 1.5$ Gyr after the big bang. With *JWST*/NIRCam, we can select galaxies at those redshifts in the rest-optical, accurately constrain their stellar masses and redshifts based on precision photometry at $1-5$ μm, provide a complete census of the galaxy population, including red galaxies previously missed by *HST*, and assess more broadly the contribution of red galaxies to the SMF. Moreover, large samples of galaxies selected with *HST* in this range and in the same parts of the sky now observed with *JWST* allow us to statistically and self-consistently compare galaxy counts and physical properties between galaxy samples obtained through the two space telescopes, providing important consistency checks as well as a comprehensive overview over the galaxy population, and the stellar mass budget of the Universe at $z \sim 4$–9.

We proceed as follows: In Section 2 we describe the *JWST* + *HST* imaging used in this work, as well as the production of the photometric catalogues, the sample selection, the spectral energy distribution (SED)-fitting procedure, and the derivation of the SMFs. We present our results in Section 3, and discuss some implications and caveats in Section 4. Finally, we summarize our findings in Section 5.

When computing distance-dependent quantities, we assume a flat cold dark matter cosmology with $\Omega_m = 0.27$, $\Omega_\Lambda = 0.73$, and $H_0 = 70$ km s$^{-1}$Mpc$^{-1}$. Fluxes and magnitudes are specified in the AB system (Oke & Gunn 1983) and our stellar mass estimates are based on a broken power-law IMF as described in Eldridge et al. (2017) based on Kroupa, Tout & Gilmore (1993) (see Section 2.4 for details). Whenever it is relevant for comparisons, we convert masses based on a Salpeter (1955) or Chabrier (2003) IMF to Kroupa et al. (1993) adopting the factors specified in Madau & Dickinson (2014).

## 2 DATA AND METHODS

### 2.1 Image reduction

In this paper, we use publicly available *JWST* imaging data over four extragalactic legacy fields, all of which had previously been observed with *HST* as part of the Cosmic Assembly Near-infrared Deep Extragalactic Legacy Survey (CANDELS; Grogin et al. 2011; Koekemoer et al. 2011) and the Great Observatories Origins Deep Survey (GOODS; Giavalisco et al. 2004a). These fields are the Extended Groth Strip (EGS, CEERS), COSMOS, the Ultra-deep Survey (UDS), and the GOODS-S.

The *JWST* imaging we use was obtained by several surveys: (1) the Cosmic Evolution Early Release Science Survey (CEERS, program ID 2079, PI Finkelstein, Bagley et al. 2023; Finkelstein et al. 2023) with additional imaging data in *F444W* from program ID 2279 (PI Naidu) and in various filters from program ID 2750 (PI Arrabal-Haro) in the EGS. (2) by the Public Release Imaging for Extragalactic Research (PRIMER, program ID 1837, PI Dunlop, Dunlop et al. in preparation) in the UDS and COSMOS fields which partially overlaps with COSMOS-Web (program ID 1727, PIs Kartaltepe & Casey, Casey et al. 2023). (3) In the GOODS-S field, we use all available NIRCam data from the second data release of the *JWST* Advanced Deep Extragalactic Survey (JADES, program ID 1180, PI Eisenstein, Eisenstein et al. 2023a) with additional imaging, partially overlapping with the JADES DR2 footprint in (a) *F182M*, *F210M*, and *F444W* from First Reionization Epoch Spectroscopically Complete Observations (FRESCO, program ID 1895, PI Oesch, Oesch et al. 2023), (b) various filters from program ID 2079 (PI Finkelstein), (c) five medium bands from the *JWST* Extragalactic Medium-band Survey (JEMS, program ID 1963, PI Williams, Williams et al. 2023), (d) six wide filters from the Parallel wide-Area Nircam Observations to Reveal And Measure the Invisible Cosmos (PANORAMIC, program ID 2514, PIs Williams and Oesch), and (e) various medium bands with exposure times reaching > 40 h from program ID 3215 (PI Eisenstein, Eisenstein et al. 2023b).

**Table 1.** 5$\sigma$ depths in all photometric filters used for each field respectively, specified in AB magnitudes; 'det-img'. refers to the stacked $F277W$ + $F356W$ + $F444W$ detection image. The last column specifies the combined survey area covered by each filter in arcmin$^2$. Note that the quoted magnitudes are `average` depths. The total survey area is *defined* by the area covered by the six NIRCam wide filters $F115W$, $F150W$, $F200W$, $F277W$, $F356W$, and $F444W$ (see Section 2.3), which is why their combined area is identical. The survey area per field is specified in Section 2.7.

| Filter | EGS (mag) | COSMOS (mag) | UDS (mag) | GS (mag) | area (arcmin$^2$) |
|---|---|---|---|---|---|
| ACS | | | | | |
| $F435W$ | 28.15 | 28.10 | 27.32 | 28.75 | 369.7 |
| $F606W$ | 28.32 | 27.97 | 28.03 | 29.01 | 424.2 |
| $F775W$ | – | – | – | 28.28 | 64.6 |
| $F814W$ | 28.17 | 27.89 | 27.96 | 28.80 | 442.6 |
| $F850LP$ | – | – | – | 28.11 | 67.3 |
| WFC3 | | | | | |
| $F105W$ | 27.93 | 27.53 | 27.57 | 28.29 | 108.1 |
| $F125W$ | 27.69 | 27.58 | 27.66 | 28.31 | 410.5 |
| $F140W$ | 27.01 | 26.98 | 26.97 | 27.02 | 336.6 |
| $F160W$ | 27.80 | 27.62 | 27.65 | 28.08 | 417.0 |
| NIRCam | | | | | |
| $F090W$ | – | 27.85 | 27.71 | 29.26 | 411.0 |
| $F115W$ | 28.71 | 27.75 | 27.75 | 29.45 | 500.8 |
| $F150W$ | 28.62 | 27.96 | 27.97 | 29.41 | 500.8 |
| $F162M$ | – | – | – | 29.73 | 8.9 |
| $F182M$ | – | – | – | 28.56 | 45.3 |
| $F200W$ | 28.86 | 28.26 | 28.19 | 29.51 | 500.8 |
| $F210M$ | – | – | – | 28.42 | 44.3 |
| $F250M$ | – | – | – | 30.02 | 9.1 |
| $F277W$ | 29.16 | 28.62 | 28.51 | 29.93 | 500.8 |
| $F300M$ | – | – | – | 30.48 | 9.1 |
| $F335M$ | – | – | – | 29.78 | 33.9 |
| $F356W$ | 29.28 | 28.84 | 28.57 | 29.92 | 500.8 |
| $F410M$ | 28.36 | 28.05 | 27.85 | 29.41 | 489.4 |
| $F430M$ | – | – | – | 28.67 | 9.6 |
| $F444W$ | 28.84 | 28.36 | 28.21 | 29.20 | 500.8 |
| $F460M$ | – | – | – | 28.38 | 9.6 |
| $F480M$ | – | – | – | 28.87 | 9.6 |
| det.-img. | 29.70 | 29.19 | 28.97 | 30.29 | 500.8 |

These *JWST* data are complemented with available *HST* imaging from all the many surveys that have covered these fields over the past two decades or so. Most important among these is the CANDELS survey (Grogin et al. 2011; Koekemoer et al. 2011), which provided optical to NIR data with *HST*/ACS and WFC3/IR. However, we use all available data in the *HST* archive over these fields in the standard ACS and WFC3/IR filters that are listed in Table 1.

All the calibrated *HST* exposures and the level-2 calibrated *JWST* NIRCam exposures were retrieved from the STScI MAST archive and were further processed with the GRIZLI software package (Brammer 2023). They were aligned to a common pixel-grid with a pixel size of 0.04 arcsec. For a basic outline of the individual reduction steps see, e.g. Valentino et al. (2023). The images used here are v7 reductions that are all publicly available from the DAWN *JWST* archive (DJA).[1]

---
[1] https://dawn-cph.github.io/dja/imaging/v7/

## 2.2 Photometric catalogues

In the following text, the generation of the photometric catalogues used in this work is described in some detail. The procedure is the same as for catalogues that have already been used in various published papers (e.g. in Atek et al. 2023; Barrufet et al. 2023 and Gottumukkala et al. 2024). The basic tool used for our photometric measurements is SOURCEEXTRACTOR (SE; Bertin & Arnouts 1996).

Starting from the GRIZLI v7.0 mosaics, we combine the provided weight files with the exposure maps and the science images to construct 'full' weight maps, including Poisson noise from the flux of sources as well as from the background.[2]

We convert those 'full' weight images to rms images as rms = $1/\sqrt{\text{weight}}$ and derive flag images. The basic idea of the flag images is to track the image locations that have reliable photometric coverage in a given filter. Therefore, we flag pixels that either have no weight (i.e. no exposure time), or that are particularly uncertain (i.e. above a given threshold value in rms).

### 2.2.1 PSF extraction and matching

In order to measure consistent colours across the different available filters, we have to take into account the wavelength dependence of the point spread function (PSF). We wish to produce a PSF-matched photometric catalogue, matching all filters to the PSF resolution in the reddest *JWST*/NIRCam wide filter, $F444W$.

We extract PSFs in each field and for all filters directly from the mosaics. First, we identify stars in each filter from a preliminary catalogue as

$$19 < \text{mag(FIL)} < 25 \text{ mag} \wedge 1.2 < f(0.35'')/f(0.16'') < 1.4, \quad (1)$$

where $f(0.35'')/f(0.16'')$ is the ratio of fluxes measured in circular apertures of radii 0.35 and 0.16 arcsec in $F444W$ and mag(FIL) is the AB-magnitude measured in the respective filter. We further exclude stars with flagged (weight = 0) pixels within a 4.04 arcsec × 4.04 arcsec cut-out around their centroid position, which usually happens if they are saturated. Finally, we look for neighbouring sources around each star, and remove those with a neighbour within 2.5 arcsec that is 2.5 mag fainter than the star or brighter.

The PSF may vary as a function of the position across the footprint of a given field. After conducting some tests, we however conclude that for our adopted aperture radius of 0.16 arcsec (see below), those differences only affect the encircled energies on the 1 per cent level or below. We therefore neglect them, and derive one PSF per field.

We use the python tool `psf.EPSFBuilder`[3] (Anderson & King 2000; Anderson 2016), which is part of the `photutils` package (Bradley et al. 2022), to derive an effective PSF from the background-subtracted cut-outs around the selected stars. We use a normalization radius of 10 pixels (0.4 arcsec), $\sigma$-clipping with a 5$\sigma$ cut, and a maximum of 10 clipping-iterations, a quartic smoothing kernel, and a maximum of 50 PSF fitting/modelling iterations. For the NIRCam short-wavelength channel (NIRCam/SW) filters, for which the full width at half-maximum (FWHM) is close to the pixel scale, we apply an oversampling of 3 and use a quadratic smoothing kernel, which yield better results. In the JADES/GOODS-S footprint, we use images sampled to a smaller 0.02 arcsec pixel scale in the NIRCam/SW filters to extract PSFs in which case we adjust the normalization radius to 20 pixels and need no oversampling to obtain accurate PSFs. We then resample those PSFs to the 0.04 arcsec pixel scale which is used for all photometric measurements.

---
[2] https://dawn-cph.github.io/dja/blog/2023/07/18/image-data-products/
[3] https://photutils.readthedocs.io/en/stable/api/photutils.psf.EPSFBuilder.html

We compute matching kernels from all the ACS and NIRCam PSFs to the NIRCam/*F444W* PSF using the software package PYPHER (Boucaud et al. 2016) with the regularization parameter set to 1e-4 and convolve each flux and rms image with the corresponding kernel to match the PSF resolution in *F444W*. Fluxes and flux uncertainties for each source (see below) are measured from those PSF-matched images.

For the WFC3 filters whose PSFs are broader than the NIR-Cam/*F444W* PSF, we follow a different procedure. First, we compute kernels matching all of them to the broadest PSF among them, WFC3/*F160W*, using pypher as described above. Then, we also produce a matching kernel from *F444W* to *F160W* and generate flux and rms images, PSF-matched to *F160W*, for all the WFC3 filters and *F444W*. In order to match the WFC3 fluxes to the PSF-resolution in *F444W*, we use the following equality:

$$\mathrm{mag}(F444W)\big|_{\mathrm{psfm.F160W}} - \mathrm{mag}(FIL)\big|_{\mathrm{psfm.F160W}}$$
$$= \mathrm{mag}(F444W)\big|_{\mathrm{psfm.F444W}} - \mathrm{mag}(FIL)\big|_{\mathrm{psfm.F444W}}, \quad (2)$$

where FIL stands for any WFC3 filter in this case. The equality simply states that the colour measured between two filters is independent of the filter to which the images are PSF-matched. Knowing both terms on the left side and the first term on the right side, we solve this equality for $\mathrm{mag}(FIL)\big|_{\mathrm{psfm.F444W}}$ to get the flux of a given WFC3 filter, PSF-matched to *F444W*. This is to avoid having to perform a deconvolution on the WFC3 images to match them to the (higher) PSF-resolution in *F444W*.

Fig. 1 illustrates the PSF extraction and matching process. On the top, we show PSFs for the NIRCam/SW filter *F200W*, extracted from the four different fields used in this work (CEERS-EGS, PRIMER-UDS, PRIMER-COSMOS, and JADES-GS) as well as a generic PSF generated through webbpsf. For the latter, we set the jitter_sigma parameter to 0.022 which has been shown to well reproduce radial profiles of NIRCam-observed stars in Morishita et al. (2024), and also does so in our case. The bottom left panel shows logarithmic radial profiles of all the PSFs shown in the top panels which agree very well, in particular at small radii ≲0.35 arcsec, including our adopted aperture radius of 0.16 arcsec (see below). In the bottom right of Fig. 1 we show the encircled energies of four different PSFs extracted from the CEERS field, divided by the encircled energies of the corresponding *F444W* PSF. We chose one PSF from *HST*/ACS (*F606W*), *HST*/WFC3 (*F125W*), NIRCam/SW (*F150W*), and NIRCam/LW (*F277W*), respectively, for illustrative purposes. Then, we match each of the four PSFs to its respective reference PSF (*F160W* for *F125W*; *F444W* for all others) by convolving it with the corresponding matching kernel and recompute the encircled energies. The resulting plot, showing them relative to the respective reference PSF (*F444W* or *F160W*), serves as an internal check of the PSF-matching procedure and demonstrates its self-consistency. For a perfect matching, the displayed ratio would be = 1 at all radii. In practice, the residuals are <1 per cent for the *F125W* (and all filters that are PSF-matched to *F160W*) and ≪1 per cent for all filters matched to *F444W*. The displayed filters are representative for the measured residuals in all filters.

### 2.2.2 Source extraction and flux measurements

The software SE (Bertin & Arnouts 1996) is used to detect sources and to perform photometric measurements. SE is run in dual image mode using an inverse-variance weighted stack of the original (unconvolved) *F277W* + *F356W* + *F444W* images as the detection image, a detection threshold of 1.3 relative to the rms image, a minimum area of 7 pixels, and a deblending parameter of 3e-4. Before detecting sources, the image is smoothed with a 3×3 pixels Gaussian filter with an FWHM of 1.5 pixels (following the same procedure as in Weaver et al. 2024). We measure fluxes on the PSF-matched images in each filter, respectively, in circular apertures with four different radii, 2.5, 4, 5, and 8.75 pixels (0.1, 0.16, 0.2, and 0.35 arcsec).

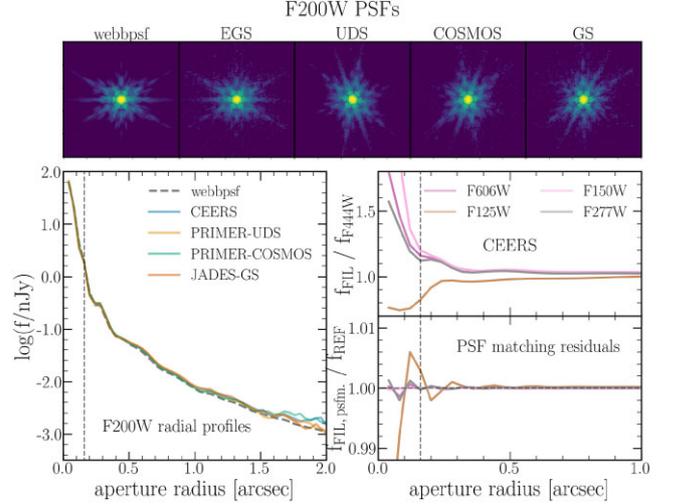

**Figure 1.** Top panels: *F200W* PSFs generated through webbpsf with jitter_sigma = 0.022 (see Morishita et al. 2024), as well as extracted from the images for all the different fields used in this work as explained in the text. All PSFs are shown in log-scale. Bottom left: logarithmic radial profiles for all the PSFs shown in the top panels, showing good agreement with each other and with webbpsf – in particular at small radii. Bottom right (upper panel): encircled energies relative to the corresponding *F444W* PSF for four different PSFs representing the four different cameras/detectors used in this work – extracted from the CEERS field as an example. Bottom right (lower panel): consistency check for the PSF-matching procedure. Each PSF is matched to its corresponding reference PSF (*F160W* for *F125W*, *F444W* for all the other filters), then encircled energies relative to *F444W* (or *F160W*) are plotted. For a perfect PSF-matching, this should yield 1 – independent of the radius. In practice, we find small residuals <1 per cent for all the filters PSF-matched to *F160W* and ≪1 per cent for all filters PSF-matched to *F444W*. The vertical dashed lines indicate our adopted aperture size of 0.16 arcsec.

We scale all fluxes to the flux measured in Kron-like apertures by SE in a PSF-matched, inverse-variance weighted stack of the NIRCam long-wavelength channel wide filters (*F277W* + *F356W* + *F444W*), where the Kron ellipse itself is inferred from the detection image using the default Kron parameters 2.5 and 3.5. Whenever the area encircled by the Kron ellipse is smaller than the area of the circular aperture, we use the circular aperture fluxes and do not apply any scaling. Next, we scale the fluxes to 'total' fluxes by measuring the fraction of the encircled energy of the Kron ellipse (or the circular aperture) on the *F444W* PSF and dividing the Kron-corrected fluxes by that fraction. Whenever the Kron ellipse is larger than the 4.04 arcsec × 4.04 arcsec PSF, we instead approximate it as a circle with radius $\sqrt{ab} \times kron\_radius$, where $a$, $b$, and $kron\_radius$ are the SE output characterizing the Kron ellipse, and infer the fraction of the encircled energy from a simple radius versus encircled energy table generated through webbpsf. Finally, we correct all fluxes for Milky Way foreground extinction using the $E(B - V)$ map from Schlafly & Finkbeiner (2011) and the extinction model from Fitzpatrick & Massa (2007) through the python package `extinction`.

### 2.2.3 RMS correction and image depths

We estimate the true rms by summing the pixels within apertures put on the flux image divided by the rms image. First, we choose 5000 random positions without any nearby objects or flagged pixels according to the flag images and the segmentation map from a preliminary SE run with low relative detection threshold (0.9) and minimum area (5 pixels). We perform measurements in circular apertures of radii 1,2,...,10 pixels, measure the scatter among the apertures of a given size, respectively, and divide it by the aperture radius times $\sqrt{\pi}$ to get the measured scatter per pixel. In theory, if the rms image perfectly describes the (random) noise and there is no correlated noise, this scatter measured on what is effectively a signal-to-noise image should be equal to one (see e.g. Whitaker et al. 2011, for further reading). In practice, it depends on the aperture size and the photometric filter, and is typically in the range 0.8–1.5. To obtain the appropriate scaling factor for a given aperture size, we linearly interpolate between the measured values for 1,2,... 10 pixel apertures. We multiply the uncertainties on all fluxes in a given filter, measured from the rms map, respectively, by this factor. Further, we apply an error floor of 5 per cent to the total errors to account for remaining systematic uncertainties and to allow for more flexibility in the SED-fitting described below.

We use a similar procedure to estimate the $5\sigma$ depths of each image, respectively, in different circular apertures, by measuring the scatter among the fluxes measured in the 5000 randomly placed apertures described above on the flux image, respectively. The resulting $5\sigma$ depths in each field and filter are listed in Table 1. Those have to be interpreted as *average* depths over the entire field. In particular, the GOODS-S field has *JWST* imaging data from various programs, leading to varying depths across the field, while the other three fields have more homogeneous coverage in most of the available filters.

We measure $0.3 - 0.5$ mag lower depths compared to Finkelstein et al. (2024) for the SW filters in CEERS, while our LW depths are consistent or slightly higher. Eisenstein et al. (2023a) list three different depths per filter, corresponding to the 'deepest', 'deep', and 'medium' parts of JADES. Our depths measured in GOODS-S lie between the 'medium' and 'deep' values for all filters, except for *F200W* where our value is slightly below the 'medium' depth. Our GOODS-S imaging however also contains imaging from Eisenstein et al. (2023b) and other programs (see Section 2.1). The former specify depths > 30 mag in all filters. As expected, we do measure comparable depths in *F250M* and *F300M* where all the imaging used in this work comes from the JADES Origins Field. Further, our depths are consistent within 0.3 mag with those published in Williams et al. (2023) for the *F430M*, *F460M*, and *F480M* imaging from JEMS, as well as with the depths specified for the PRIMER-COSMOS, -UDS, and the JADES GOODS-S field in Donnan et al. (2024). Differences with respect to published depths can be attributed to differences in the data reduction, the methodology applied to measure them (e.g. the aperture size used or whether they are measured on PSF-matched images), and the averaging over different parts of a given field with varying depth.

We run the SED fitting code `eazy` (Brammer, van Dokkum & Coppi 2008) on the final total fluxes and errors, using the `blue_sfhz` template set.[4] This consists of 13 templates gener-ated through the FLEXIBLE STELLAR POPULATION SYNTHESIS code (Conroy, Gunn & White 2009; Conroy & Gunn 2010), with redshift-dependent SFHs and physical properties. To account for extreme emission lines observed at high redshifts, the best fit to an NIRSpec spectrum of a galaxy with strong emission lines at $z \sim 8.5$ from Carnall et al. (2023a) is added as a 14th template. We use the zero-point optimization in `eazy`, to correct for possible remaining calibration offsets based on the template fitting residuals. Performing three iterations we find small corrections of typically $1 - 4$ per cent for all the *JWST* and *HST* filters, with the exception of correction factors up to 8 per cent for some *JWST* filters in GOODS-S, and of up to 13 per cent for the *HST B* band (F435W). Finally, we allow the best-fitting redshift to be in the range $z_{\rm phot,\,eazy} \in (0.01, 20)$.

[4] https://github.com/gbrammer/eazy-photoz/tree/master/templates/sfhz

### 2.2.4 Flags

We mask bright saturated stars, together with their diffraction spikes as well as noisy regions – mostly along the edges of the images – by hand, to avoid having to deal with objects that are contaminated by stellar light and or with substantially enhanced noise. This step is performed using the software tool `DS9`. The areas that are to be flagged are marked using polygon-shaped regions on the detection image. We convert the resulting region file into a binary pixel-by-pixel mask using the python package `Polygon3` and include the masked regions as 1-valued pixels in our flag images in all filters. In order to flag corresponding sources generously, we apply a binary dilation[5] 10 times to each flag image. We then define binary flags flag_FIL for every filter FIL. For each object in the catalogue, if its isophotal footprint according to the segmentation map from SE overlaps with one or more flagged pixels in the dilated flag image, we set the corresponding flag_FIL = 1 and the total flux and error to $-99$ – so that the corresponding measurements are ignored in the subsequent SED-fitting runs.

In order to also flag less bright and unsaturated stars, we define a stellar_flag, for which we use two different criteria: first, we identify bright point sources (mag($F444W$) < 24.5) through their flux ratio $1.2 < F(0.35'')/F(0.16'') < 1.4$, in analogy to equation (1). Secondly, we match our catalogues to sources in the *Gaia* DR3 catalogue (Gaia Collaboration 2016, 2023) with non-zero proper motion, using a large search radius of 2.5 arcsec to include diffraction spikes and nearby contaminated sources around non-saturated stars. In our full catalogues over all four fields combined, this flags 0.5 per cent of the sources.

We further define a huge_flag with which we flag sources with very large spatial extent ($R_{50}(F444W) > 50$ pix) where $R_{50}(F444W)$ is the half-light radius measured by SE in *F444w* or unreasonably large Kron radii ($\sqrt{(ab)}$ kron_radius > 150 pix) where a, b, and kron_radius are adopted from the SE output. The former criterion flags some extended foreground galaxies with unproblematic photometry but those are irrelevant for this work as we only consider galaxies at $z \gtrsim 3$ (see below). In total, only 0.1 per cent of all sources are flagged as 'huge'.

Flagging sources with $0 < R_{50}(F444W) < 1.2$ pixels and mag($F444W$) < 28.5 (approximately corresponding to the $5\sigma$ depths of the images, cf. Table 1) reliably identifies some remaining spurious detections and hot pixels which we specify as a junk_flag. This flags another 0.5 per cent of all objects in the catalogues.

[5] https://docs.scipy.org/doc/scipy/reference/generated/scipy.ndimage.binary_dilation.html

Finally, we flag sources with an S/N ratio < 3 in all of *F115W*, *F150W, F200W, F277W, F356W,* and *F444W* (sn_flag, 2.3 per cent of all sources), which is largely but not entirely superseded by the S/N cut applied in Section 2.3, as well as sources for which eazy returns a best-fitting $\chi^2$ value of $-1$ (nofit_flag). The latter affects as many as 10.7 per cent of all sources, the vast majority of which have no best-fitting redshift ($z_{\text{phot, eazy}} = -1$). In most cases, this is due to missing filters, since we set up eazy to require at least five filters to provide a fit. This is entirely superseded by requiring all six NIRCam wide filters from *F115W* to *F444W* in Section 2.3. The nofit_flag however also includes sources with $z_{\text{phot, eazy}} \approx 20$ where the best-fitting solution converged to the edge of the prior (0.6 per cent).

### 2.3 Sample selection

Given *HST* cannot probe the rest-frame optical at $z \gtrsim 3$, we focus our present efforts on selecting sources in the range $z \sim 4 - 9$, exploiting the unique near-infrared capabilities of *JWST* which gives us access to the rest-frame optical at these redshifts. We use eazy to broadly select galaxies at $z \gtrsim 3$, in order to reduce the sample size for the subsequent more detailed SED-fitting. To this end, we apply the following selection criteria.

(i) The best-fitting eazy redshift is > 3.

(ii) P($z_{\text{phot, eazy}}$ > 2.5) > 0.8 where P(z) is the eazy posterior redshift distribution.

(iii) (S/N)$_{\text{det}}$ > 10 where (S/N)$_{\text{det}}$ is the signal-to-noise ratio measured in the stacked *F277W* + *F356W* + *F444W* detection image. For a flat SED, this approximately corresponds to a S/N ratio of 5.8 in each of the three filters. We chose this threshold as a trade-off between pushing to low stellar masses at high redshifts while ensuring that the inferred stellar masses and redshifts are sufficiently reliable. At (S/N)$_{\text{det}}$ > 10 the median uncertainty in the photometric redshift inferred by bagpipes (see Section 2.4) is well below 1.

(iv) Available data in the NIRCam wide filters *F115W, F150W, F200W, F277W, F356W,* and *F444W* (i.e. flag_FIL = 0 in all the mentioned filters, see Section 2.2.4).

(v) None of the flags described in Section 2.2.4 is set (i.e. stellar_flag = huge_flag = junk_flag = sn_flag = nofit_flag = 0).

We visually inspect all of the selected sources in the detection image and remove remaining spurious detections (1437 in total from all four fields) – mostly in the enhanced background noise around bright extended foreground sources and diffraction spikes that are not captured by the corresponding flags (Section 2.2.4).

This selection yields 45 266 objects in total, of which 9165 are in CEERS, 9754 in PRIMER-COSMOS, 12 822 in PRIMER-UDS, and 13 525 in JADES-GS. In the following text, we assume that in the range $3.5 < z < 9.5$, this sample is neither systematically missing galaxies, nor is it substantially contaminated by low-redshift interlopers (see also Fig. 2).

### 2.4 Bagpipes SED fitting

While eazy in principle provides redshifts as well as stellar masses for each source, we run fits with the Bayesian Analysis of Galaxies for Physical Inference and Parameter EStimation tool (bagpipes; Carnall et al. 2018) to obtain the final redshifts and physical properties for our selected galaxies. There are two main reasons for this: first, eazy stellar masses are based on combining the fitted templates which may not always be physically meaningful. Secondly, we wish to self-consistently infer all the relevant parameters within the Bayesian framework of bagpipes because this allows us to sample SMFs from the posterior distributions, yielding unbiased results and accurate uncertainties (see Section 2.7).

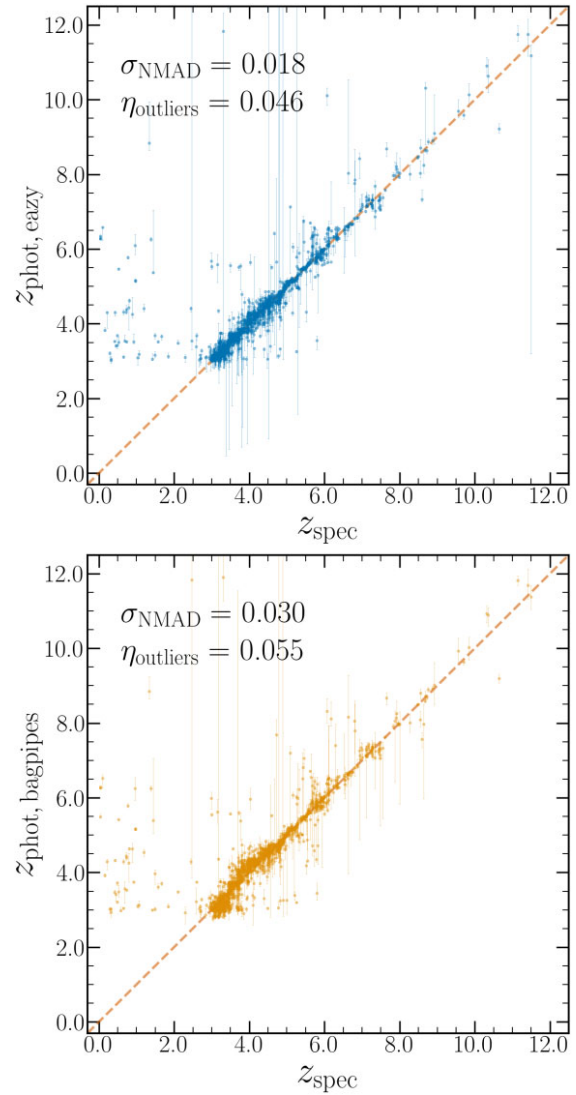

**Figure 2.** Comparison between photometric and spectroscopic redshifts. Spectroscopic redshifts from the literature are plotted against photometric redshifts obtained through eazy in the top panel and through bagpipes in the bottom panel. Both codes perform well at $z \gtrsim 3$ with eazy showing a smaller normalized median absolute deviation $\sigma_{\text{NMAD}}$ and outlier fraction $\eta_{\text{outliers}}$, defined as the fraction of sources with $\Delta z/(1 + z_{\text{spec}}) > 0.15$.

We do however assume that eazy correctly identifies $z \gtrsim 3$ galaxies, and constrain the redshift to the range z =(2.5,20) with a uniform prior – allowing for some flexibility at the lower end but avoiding cases where bagpipes would prefer a low-redshift solution since we do not take the opposite case into account where eazy would prefer a lower redshift and bagpipes a redshift $z \gtrsim 3$. Allowing for lower redshifts in bagpipes would therefore bias our results. For the vast majority of the sources, the redshift inferred by bagpipes does not converge towards the edge of the prior but instead, a plausible high-$z$ solution is found which is typically consistent with the solution found by eazy.

We use a delayed-$\tau$ model for the SFH, with broad uniform priors in age (i.e. the time since star formation began) ranging from 0.01 to 5 Gyr as well as in the logarithm of $\tau$, $\log(\tau) \in (0.1,10)$. Note that bagpipes does not allow star formation before the big bang, so the upper limit on the age varies as a function of the observed redshift. We discuss the possible systematic effects on the inferred stellar masses implied by the choice of a specific SFH-model in Section 4.3.

We use the BPASS-v2.2.1 stellar population models (Stanway & Eldridge 2018) assuming a broken power-law IMF with slopes of $\alpha_1 = -1.3$ from $0.1 - 0.5 \, M_\odot$ and $\alpha_2 = -2.35$ from $0.5 - 300 \, M_\odot$ as described in Eldridge et al. (2017) based on Kroupa et al. (1993). Further, we use a Calzetti dust attenuation curve (Calzetti et al. 2000) with a uniform prior on the extinction parameter $A_V \in (0, 5)$ as well as on the stellar metallicity $Z \in (0.1, 1) \, Z_\odot$.

The nebular emission in bagpipes is modelled using CLOUDY (Ferland et al. 2017). We compute a grid of the nebular emission models from the BPASS-v2.2.1 stellar population models as described in Carnall et al. (2018) but extending the allowed range of ionization parameters in bagpipes to $\log(U) \in (-4, -1)$ with a uniform prior to account for the strong rest-frame optical emission lines that are observed in early galaxies.

This leaves us with seven free parameters in the SED-fitting: two for the delayed-$\tau$ SFH, redshift, $A_V$, Z, $\log(U)$, and a normalization which is expressed as the total mass formed, i.e. the integral of the SFH in the bagpipes set-up. The latter is allowed to be in the range $\log(M_{\rm formed}/M_\odot) \in (5, 13)$, with a uniform prior.

In addition to the standard physical properties estimated by bagpipes, we measure the UV-slope $\beta$ of each source by performing a simple linear regression fit to the best-fitting SED in the wavelength range 1350 Å $< \lambda <$ 2800 Å. We specifically use this to split our sample into UV-red ($\beta > -1.2$) and UV-blue ($\beta < -1.2$) objects in Section 3.1.2.

While visually inspecting the sample galaxies in the detection image, we identified some cases with contaminated Kron ellipses. Typically, if a faint compact source has a nearby bright source in the foreground, its Kron ellipse is extended towards the neighbouring source and contaminated by its light. As a result, all fluxes, and thus the stellar masses, will be boosted by the same factor (see Section 2.2.2). For those visually identified objects, we rerun bagpipes based on the fluxes measured in circular apertures with a radius of 0.25 arcsec, scaled to total fluxes based on the encircled energy on the F444W PSF. This affects <1 per cent of all objects.

In Fig. 2, we compare our two photometric redshift estimates (from bagpipes and eazy) to spectroscopic redshifts from the literature. In total, there are 1851 sources with spectroscopic redshifts in our sample, 256 in the CEERS field, 197 in PRIMER-COSMOS, 386 in PRIMER-UDS, and 1012 in GOODS-S. 739 of those come from publicly available NIRSpec spectra (programs with IDs 2750, PI Arrabal Haro, 1345, PI Finkelstein, e.g. Arrabal Haro et al. 2023, Fujimoto et al. 2023, 2198, PI Barrufet, Barrufet et al. 2024, 1210 and 1286, PI Luetzgendorf, 6541, PI Egami, 3215, PI Eisenstein, Eisenstein et al. 2023b, 2565, PI Glazebrook, 1180, PI Eisenstein, and 4233, PI De Graaff, e.g. de Graaff et al. 2024, Wang et al. 2024b). All those spectra were reduced using msaexp and the redshifts were taken from the DAWN *JWST* Archive (DJA).[6] The remaining 1112 redshifts come from various ground-based surveys, including the MOSFIRE Deep Evolution Field (MOSDEF; Kriek et al. 2015), the DEEP2 Galaxy Redshift Survey (Newman et al. 2013), and the VANDELS survey (Garilli et al. 2021). We assess the performance of the two photometric redshift estimators using the statistic $\sigma_{\rm NMAD}$ (Hoaglin, Mosteller & Tukey 1983; Brammer et al. 2008), defined as 1.4826× the median absolute deviation of the normalized redshift differences $\Delta z/(1 + z_{\rm spec})$ and the outlier fraction $\eta_{\rm outliers}$, defined as the fraction of sources with $\Delta z/(1 + z_{\rm spec}) > 0.15$. According to these two values, indicated in each panel in Fig. 2, eazy performs slightly better, showing $\sigma_{\rm NMAD} = 0.018$ and $\eta_{\rm outliers} = 0.050$, while for bagpipes we find $\sigma_{\rm NMAD} = 0.030$ and $\eta_{\rm outliers} = 0.057$. Comparing the two photometric redshift estimates to each other, without limiting the sample to sources with spectroscopic redshifts, we find $\sigma_{\rm NMAD} = 0.024$ and an outlier fraction of $\eta_{\rm outliers} = 0.057$. This relatively good agreement is partly related to the redshift prior applied in bagpipes, where we require $z_{\rm bagpipes} > 2.5$, avoiding some outliers where bagpipes would prefer a lower redshift, as explained above. At the same time, there are some degeneracies where two or more plausible redshift solutions are found by both eazy and bagpipes, resulting in multiply peaked probability distribution functions. If the two codes prefer a different solution, this contributes to the outlier fraction, but it is later mitigated by our sampling of the physical properties from the bagpipes posterior as described in Section 2.7.4.

We conclude that both SED-fitting codes yield satisfactory redshifts, and for the reasons outlined above, we use the redshifts and physical properties obtained through bagpipes in all subsequent plots and analyses.

In Fig. 3, we show the redshifts and stellar masses from our bagpipes runs in all four fields separately. It is apparent that GDS is the deepest of the four fields, allowing us to probe significantly lower masses at any given redshift compared to the other three fields. Little red dots (LRDs; Matthee et al. 2024) and red compact objects defined according to Section 2.5 are shown as filled and empty diamonds. They typically have high inferred masses $> 10^9 M_\odot$, and constitute an outlier population in the $\log(M_*) - z$ diagrams at $z \sim 7 - 8$.

## 2.5 Active galactic nucleus contamination

Accounting for light coming from active galactic nuclei (AGNs) rather than stars when computing the SMF has been discussed by e.g. Grazian et al. (2015), Davidzon et al. (2017), and Weaver et al. (2023). Since bagpipes models the full SED based on stellar populations, any light coming from an AGN may bias the mass towards higher values. Removing spectroscopically confirmed or hard X-ray-detected AGNs from their sample in GOODS-S, Grazian et al. (2015) find a negligible effect on their SMFs at $z \sim 4 - 7$ for $M_* < 10^{11} \, M_\odot$. Davidzon et al. (2017) argue based on Hainline et al. (2012) and Marsan et al. (2017) that for massive galaxies at $z > 3$, disregarding the effect of AGNs in the SED-fitting may cause biases in the estimated stellar mass of 0.1–0.3 dex. Further, the AGN fraction is estimated as a function of UV absolute magnitude $M_{UV}$ at $z \sim 4$ in Bowler et al. (2021). They find a fraction close to 0 for $M_{UV} > -22$. In our sample, only 48 galaxies (0.16 per cent) have brighter UV magnitudes $M_{UV} < -22$. More broadly investigating the possible impact of UV-bright AGNs on our SMFs is beyond the scope of this work, and we subsequently assume it to be negligible. We refer the reader to Weaver et al. (e.g. 2023) for a more detailed discussion of the effect of AGN on SMFs.

However, various authors have recently discovered and discussed a surprising abundance of extremely red and compact sources at high redshift, named LRDs. While these were identified originally through broad H$\alpha$ emission (Matthee et al. 2024), they were also selected through simple, red colour cuts in NIRCam bands (e.g. Labbé et al. 2023b; Barro et al. 2024).

While there is no consistent definition of LRDs in the literature, their SEDs typically show a red continuum in the rest-frame optical and blue colours in the rest-frame UV (e.g. Furtak et al. 2024). Since bagpipes does not include an AGN-component and models dust as a simple screen attenuating all of the light coming from the galaxy, it typically cannot reproduce the SEDs of LRDs. Instead, it fits those objects as dusty star-forming galaxies with an extremely red continuum throughout and typically estimates a high, and most likely wrong, stellar mass for those sources. An example of this is shown in the top panel of Fig. B1.

---

[6] https://dawn-cph.github.io/dja/spectroscopy/nirspec/

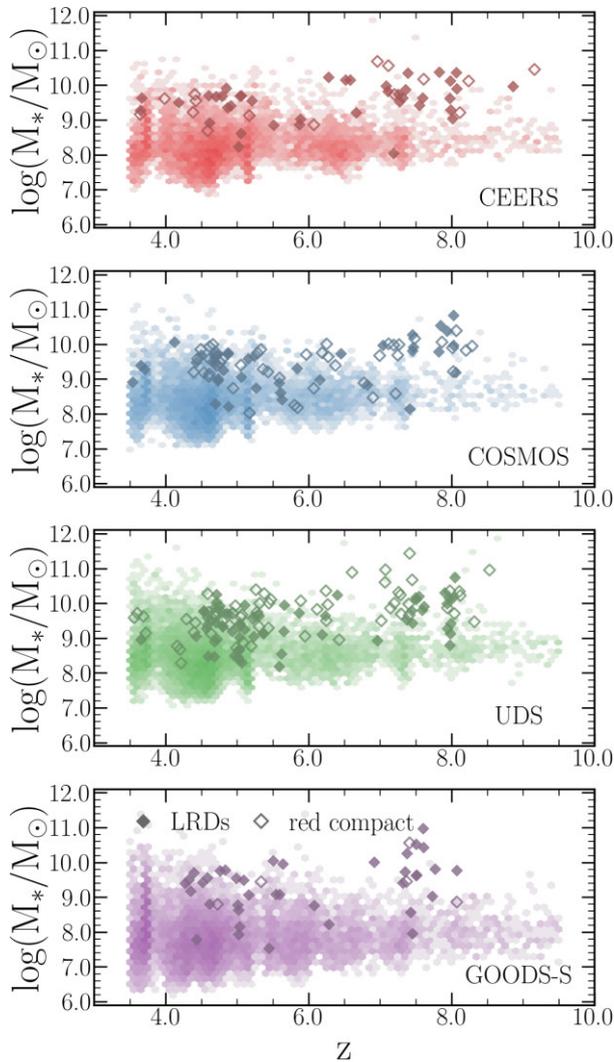

**Figure 3.** log($M_*$) − $z$ distribution of all our sample galaxies (3.5 < $z$ < 9.5), shown in each of the four fields studied in this work separately. The filled diamonds represent LRDs, and the empty diamonds represent red compact objects as defined in Section 2.5.

Labbe et al. (2023a) defined colour-cuts (red in the rest-optical, blue in the rest-UV) and a compactness criterion based on NIRCam filters to select LRDs. They complemented NIRCam photometry from Ultradeep NIRCam and NIRSpec ObserVations before the Epoch of Reionization (UNCOVER, Bezanson et al. 2022) with ALMA data and found that the selected sources whose light profiles are dominated by a PSF-component only have upper limits from ALMA, providing evidence in favour of an AGN interpretation. This was confirmed through follow-up NIRSpec spectroscopy in Greene et al. (2024), finding 14 out of 17 sources selected with the above criteria to be AGNs [the latter three being cool brown dwarf stars (Burgasser et al. 2024)].

While there is still some debate in the literature about the various selection criteria of LRD AGNs and their true nature (e.g. Pérez-González et al. 2024), in the following text, we will use the colour selection of Labbe et al. (2023a) to identify likely AGNs (and brown dwarfs) from our sample.

The selection cuts presented in Labbe et al. (2023a) are S/N(F444W) > 14 and mag(F444W) < 27.7 combined with (red1 ∨ red2) ∧ compact where

$$red1 = (mag(F115W) - mag(F150W) < 0.8) \wedge$$
$$(mag(F200W) - mag(F277W) > 0.7) \wedge$$
$$(mag(F200W) - mag(F356W) > 1.0)$$

and

$$red2 = (mag(F150W) - mag(F200W) < 0.8) \wedge$$
$$(mag(F277W) - mag(F356W) > 0.7) \wedge$$
$$(mag(F277W) - mag(F444W) > 1.0)$$

and

$$compact = f(0.2'')/f(0.1'') < 1.7 \qquad (3)$$

and f(0.2″) and f(0.1″) are the fluxes measured in 0.2″ and 0.1″ radius apertures in *F444W*, respectively. Note that Labbe et al. (2023a) use these cuts to select a parent sample. For each selected source they then fit the light profile with a PSF- and a Sérsic component. Only the objects whose fits are dominated by the PSF-component are considered reliable AGN-candidates. In their work, this is the case for 26/40 galaxies selected according to equation (3). The remaining objects, while not dominated by a point source, may still have a significant AGN-contribution to their total light.

Both red1 and red2 are a combination of red colours in the rest-frame optical and a blue colour in the rest-UV. Among the compact sources with red colours in the rest-optical, there are some that completely drop out of the relevant NIRCam/SW filters that probe the rest-UV (*F115W*, *F150W*, *F200W*). Those sources may have some UV-emission and a blue UV-continuum which is simply too faint to be detected in currently available NIRCam imaging. We therefore identify two sets of sources. The first set consists of confident LRDs, satisfying all the above selection cuts and having S/N-ratios > 3 in all the filters required to infer the relevant colours. For the second set of sources, we only apply the colour cuts in the rest-frame optical, i.e. those involving *F200W*, *F277W*, and *F356W* in red1 and *F277W*, *F356W*, and *F444W* in red2, and do not require anything regarding the S/N in the rest-UV. This second selection will then include red sources with < 3σ (non-)detections in the rest-UV that do, however, have a compact morphology, making them LRD-candidates. We refer to those as 'red compact' sources.

We identify 183 confident LRDs (38 in JADES-GS, 40 in CEERS, 40 in PRIMER-COSMOS, and 65 in PRIMER-UDS), and 138 red compact sources (8 in JADES-GS, 18 in CEERS, 46 in PRIMER-COSMOS, and 66 in PRIMER-UDS), totalling 318 sources. This is roughly consistent with Kokorev et al. (2024) who find 260 LRDs over 340 arcmin² of NIRCam imaging in the same fields studied here, and applying almost the same colour-cuts, but dealing slightly differently with UV non-detections.

Throughout this work, we show our results with and without the confident LRDs and/or the red compact objects, whenever their inclusion is relevant, or we highlight those populations specifically. In Section 4.4, we further discuss sources showing extreme masses and we show some example SEDs in Appendix B.

For our fiducial sample, we remove the confident LRDs. Further restricting the sample to sources with $3.5 < z_{\text{phot, bagpipes}} < 9.5$, we are left with a sample size of 30 631 galaxies.

### 2.6 Matching to an *HST*-based LBG sample

Bouwens et al. (2015) (B15 hereafter) compiled a sample of $\sim 10,000$ LBGs at $z \sim 4$ to $z \sim 10$ from *HST* Legacy Fields using colour–colour criteria and dropouts to identify galaxies in redshift bins ranging from $z \sim 4$–10. Their sample includes 4186 sources in the CANDELS-EGS, -UDS, -COSMOS, and GOODS-S, 2065 of which are detected in our catalogues and covered by all six required NIRCam wide filters (Section 2.3).

289 of the 2065 detected objects (14 per cent) are not in our $z \gtrsim 3$ sample according to the selection criteria in Section 2.3. Specifically, two sources do not pass the signal-to-noise threshold, i.e. they have $(S/N)_{\text{det}} < 10$; 260 sources have $z_{\text{phot, eazy}} < 3$; 12 sources do have a best-fitting redshift > 3 but do not satisfy $P(z_{\text{phot, eazy}} > 2.5) > 0.8$, and the remaining 10 sources are flagged as stars (or contaminated by stellar light) according to Section 2.2.4.

This leaves us with 1780 sources which are part of the B15 sample and of our $z \gtrsim 3$ sample, four of which are selected as LRDs according to Section 2.5. Since none of the points listed above induces any bias on the physical properties of those galaxies, we can subsequently perform a statistically meaningful comparison between their physical properties as inferred using all available *JWST* + *HST* imaging data and the physical properties of all galaxies in our $z \gtrsim 3$ sample. This will allow us to self-consistently investigate the differences between the physical properties of *HST*-based rest-frame UV detected and colour–colour selected LBGs and *JWST*-based rest-frame optically detected and photo-z selected galaxies. The results of this comparison will be shown in Section 3.1.

### 2.7 Inferring SMFs

One of the key goals of this work is the measurement of galaxy SMFs at $z > 3$. SMFs are inferred by counting the number of sources in a given bin of redshift and stellar mass and dividing by the survey volume. There are several sources of incompleteness (Sections 2.7.1 and 2.7.2), uncertainty (Section 2.7.4), and the Eddington bias (Section 2.7.5) that have to be taken into account when performing this measurement which are discussed below.

Formally, the SMF in a given mass bin i ($M_{i,\text{min}} < M_* < M_{i,\text{max}}$) and redshift bin j ($z_{j,\text{min}} < z < z_{j,\text{max}}$) can be written as

$$\Phi_{i,j}(M)[\text{dex}^{-1}\,\text{Mpc}^{-3}] = \frac{1}{b_i} \sum_q \frac{W_{i,j}(q)}{V_{\text{max},j}(q)\,C(q)}, \quad (4)$$

where the index q iterates through all objects in the sample and the window function $W_{i,j}(q)$ equals 1 if a given object falls in the ith mass-bin and the jth redshift bin, and 0 otherwise. $b_i$ is the width of the ith mass bin in dex, $C(q)$ is a magnitude-dependent completeness factor (Section 2.7.1), and $V_{\text{max},j}(q)$ is the maximum comoving volume in which source q could be observed within redshift bin j (Schmidt 1968, see Section 2.7.3).

To estimate our survey volume, we start by counting un-flagged pixels in the mosaics according to the selection criteria in 2.3. I.e., we count pixels which are not flagged in *F444W*, *F356W*, *F277W*, *F200W*, *F150W*, and *F115W*.

We find the following survey areas for each of the four fields: 82.0 arcmin$^2$ for CEERS, 127.1 arcmin$^2$ for PRIMER-COSMOS, 224.4 arcmin$^2$ for PRIMER-UDS, and 67.3 arcmin$^2$ for JADES-GS.

This yields a total survey area of $\sim 500\,\text{arcmin}^2$ or $\sim 0.14\,\text{deg}^2$. Note that not all photometric filters used in this work cover the entire survey area (see Table 1).

Subsequently, we split our sample into six redshift bins, centred at $z \sim 4, 5, 6, 7, 8,$ and 9, and with a bin size of $\Delta z = 1$ as well as into equally sized, 0.5 dex wide mass-bins ranging from $\log(M_*/\text{M}_\odot) = 7 - 12$.

#### 2.7.1 Photometric detection completeness

To assess the detection completeness of our catalogues, we use the GALAXY SURVEY COMPLETENESS ALGORITHM 2 (GLACIAR2) software (Leethochawalit et al. 2022), which builds on its predecessor GLACIAR (Carrasco et al. 2018). This software injects modelled galaxies into real images and then runs SE to obtain statistics on the fraction of recovered sources as a function of their input and output magnitude. We start by choosing a representative $1.5' \times 1.5'$ cut-out in each field (CEERS, PRIMER-UDS, PRIMER-COSMOS, and JADES-GS) that does not contain any masked stars (Section 2.2.4). On this cut-out, we run GLACIAR2 with the schechter_flat shape of the input luminosity function, performing 10 iterations with 500 galaxies per iteration, but only injecting 100 galaxies at a time to avoid overcrowding. Sources are injected in 34 magnitude bins ranging from 22.5 to 30.5 AB.

In this work, we only consider the detection completeness as a function of magnitude and therefore simulate all galaxies at a fixed redshift of 6 and with a flat SED ($\beta = -2$). We run GLACIAR2 on the three filters *F277W*, *F356W*, and *F444W* which together form the stacked detection image (see Section 2.2). The injected galaxies follow a Gaussian distribution in the logarithm of their sizes, centred at $R_{\text{eff}} = 0.8$ kpc which is inferred from the measured half-light radii of the galaxies in our sample. They have Sérsic disc profiles with 25 per cent of the galaxies having Sérsic indices of 1 and 2, respectively, and 50 per cent having a Sérsic index of 1.5. We further use five equally sized bins in inclination and eccentricity, respectively.

Given this set-up, GLACIAR2 returns a matrix where the fraction of recovered sources is specified as a function of both input and measured output magnitude. This allows us to account for the fact that galaxies scatter between input and output magnitude bins. The completeness is then given by the fraction of recovered sources that also have S/N > 10 in the detection image. Typical completeness factors at mag(det-img.) < 28.5, i.e. above the 5$\sigma$ depths of all fields (see Table 1), range from 0.75 in JADES-GS to 0.91 in PRIMER-UDS.

#### 2.7.2 Mass completeness

To avoid large corrections, we wish to only show SMFs in a regime where our sample is mostly complete. To this end we chose a rather conservative S/N threshold of 10 in the detection image in Section 2.3. In addition, we now derive 80 per cent mass completeness limits per field in each redshift bin following a similar approach as, e.g. Pozzetti et al. (2010). The mass limit depends on the mass-to-light ratio distribution of the galaxy population. For each source, we compute the hypothetical minimum mass $M_{*,hyp}$ at which it would still be included in our catalogue. Thus, $M_{*,hyp} = M_*/r_{S/N}$, where $r_{S/N} = (S/N)_{\text{det}}/(S/N)_{\text{thresh}}$, $(S/N)_{\text{det}}$ is the signal-to-noise ratio in the detection image and $(S/N)_{\text{thresh}} = 10$.

The 80th percentile of the distribution of all $M_{*,hyp}$ in a given redshift bin then represents the mass above which 80 per cent of all sources are observed above our signal-to-noise threshold, i.e. with $(S/N)_{det} > 10$. We round the inferred mass up to the next mass bin-edge to obtain our completeness limit. This is done for each field separately as different fields have different depths. In the following text, we only show SMFs in mass-bins where the completeness is expected to be > 80 per cent. The adopted completeness limits range from $\log(M_{*,lim}/M_\odot) = 7.5$ in GOODS-S at $z \sim 4$ to $\log(M_{*,lim}/M_\odot) = 9$ in PRIMER-UDS, -COSMOS, and CEERS at $z \sim 9$.

### 2.7.3 $V/V_{max}$

To account for mass incompleteness *within* a given redshift bin j, we apply a $V/V_{max}$ correction (Schmidt 1968). For each source q in the sample, we derive the maximum redshift $z_{max}(q)$ at which it would still be observed above the signal-to-noise threshold $(S/N)_{thresh} = 10$, using the following equation

$$\frac{d_{L,max}}{1+z_{max}} = \frac{(S/N)_{det}}{(S/N)_{thresh}} \frac{d_{L,obs}}{1+z_{obs}}, \quad (5)$$

where $z_{obs}$ is the observed redshift, and $d_{L,max}$ and $d_{L,obs}$ are the luminosity distances corresponding to $z_{max}$ and $z_{obs}$, respectively. Both $z_{max}$ and $d_{L,max}$ are unknown but related through cosmological equations, so we can solve for $z_{max}$. If $z_{max} > z_{max,bin}$ where $z_{max,bin}$ is the maximum redshift within a given redshift bin, $V_{max,j}(q) = V_j$, where $V_j$ is the comoving volume of the jth redshift bin. Else, $V_{max,j}(q) < V_j$, and dividing by $V_{max,j}(q)$ in equation (4) corresponds to assigning a weight to source q as $\Delta z/(z_{max} - z_{min,bin})$ where $\Delta z = 1$ is the width of the redshift bin. This correction only affects a small number of the faintest sources in our sample.

### 2.7.4 Uncertainties

There are three contributions to the uncertainty of the SMF measurement: the Poisson uncertainty on the number count in a given stellar mass and redshift bin, the uncertainty in stellar mass and redshift – which are a result of uncertainties in the photometry and the SED-fitting procedure – and cosmic variance.

We estimate the Poisson uncertainty using the frequentist central confidence interval (see Maxwell 2011).

In order to take uncertainties in both the photometry and the SED-fitting and the associated uncertainties in $M_*$ and z into account, we use the Bayesian nature of `bagpipes` and sample our SMFs from its posterior distributions. Specifically, we sample values of $M_*$, z, and $\beta$ for each galaxy 1000 times from the respective posterior distributions and compute 1000 corresponding SMFs (for all, as well as for 'UV-blue' and 'UV-red' sources separately, see Section 3.2.3). In each stellar mass and redshift bin, we use the median number count among the 1000 sampled values as our final measurement of the number density and the scatter as a contribution to its uncertainty.

Finally, we estimate cosmic variance following the methodology presented in Moster et al. (2011) through the publicly available python package `cosmic-variance`[7], taking field geometry into account, and using equation (7) in Moster et al. (2011) to combine the cosmic variances inferred for each field to a total cosmic variance. The inferred values strongly increase with mass and redshift. At $z \sim 4$ they range from an uncertainty of 10 per cent for the $8.5 < \log(M_*/M_\odot) < 9$ bin to ∼37 per cent at the high-mass end, while at $z \sim 9$, the equivalent values are 70 per cent and 191 per cent. At the lowest masses shown in each redshift bin, the contribution of cosmic variance slightly increases as the corresponding SMF values are not inferred from all four fields, but only from CEERS and/or JADES-GS which are sufficiently deep to probe those masses (see Section 2.7.2). In Appendix A, we present SMFs for each field separately. The scatter among them provides a constraint on the cosmic variance, and it is consistent with or smaller than the values inferred based on Moster et al. (2011).

---

[7] https://pypi.org/project/cosmic-variance/

The three uncertainties are added in quadrature to obtain a total uncertainty.

### 2.7.5 Schechter fitting parameters and Eddington bias

Having derived the binned SMFs, we fit a single Schechter function (Schechter 1976), characterized by a power-law slope $\alpha$, a characteristic mass $M^*$, above which the function drops exponentially, and a normalization $\Phi^*$. Specifically, we perform the fitting to each of the 1000 SMF realizations drawn from the `bagpipes` posterior distributions (for UV-red, UV-blue and all galaxies respectively, Section 3.2.3). We use the inferred median values for $\alpha$, $M^*$, and $\Phi^*$ as our final Schechter fitting parameters and the scatter among the values derived in the 1000 respective fits as the uncertainty of the fitting parameters.

As has been known for a long time, the number count of galaxies in a given mass and redshift bin is an overestimate of the true number count due to the so-called Eddington bias (Eddington 1913). Because of the steepness of the SMF, the number of lower mass galaxies scattering into bins of higher mass is always larger than vice versa, therefore biasing the inferred SMF towards higher masses. Various approaches to account for this bias have been discussed in the literature (e.g. Ilbert et al. 2013; Grazian et al. 2015; Adams et al. 2021; McLeod et al. 2021).

We follow the approach outlined in Ilbert et al. (2013) and convolve our Schechter function by the uncertainty in $M_*$ before fitting $\alpha$, $M^*$, and $\Phi^*$. Instead of using an analytic approximation to characterize the uncertainty in $M_*$, we construct a numerical uncertainty distribution in $\log(M_*)$ in each redshift bin based on the median `bagpipes` posterior distribution. Those median distributions turn out to be quite well characterized by normal distributions with standard deviations $\sigma_{M_*} = 0.13, 0.17, 0.19, 0.20, 0.22$ and $0.25$ in our redshift bins centred at $z \sim$4, 5, 6, 7, 8, and 9 with the inferred distributions being slightly skewed towards lower masses. If we refer to the inferred median uncertainty distribution as U and the Schechter function as $\Phi$, the convolved Schechter function $\Phi_{conv}$ reads

$$\Phi_{conv}(\log M_*) = \int_{-\infty}^{\infty} \Phi(x) U(\log M_* - x) dx \quad (6)$$

All Schechter fitting parameters shown in Table 4 are obtained from fitting $\Phi_{conv}$ to our SMF data points.

## 3 RESULTS

### 3.1 Red sources with and without *JWST*/NIRCam

#### 3.1.1 $M_*$-$\beta$ distribution

One of the key questions in early galaxy science was whether the *HST*-based source selections at $z > 3$ might be missing a large fraction of red, obscured sources. Due to *JWST*'s NIR sensitivity, this question can now be answered. In Fig. 4, we show the distributions of stellar mass $M_*$ and UV-slope $\beta$ as estimated through `bagpipes` (Section 2.4) in five different redshift bins centred at $z = 4, 5,$ and 6, and a combined bin $z \sim 7 - 8$ ($6.5 < z < 8.5$), for all the sources in our sample and for the subset of sources that are matched with the B15 sample.

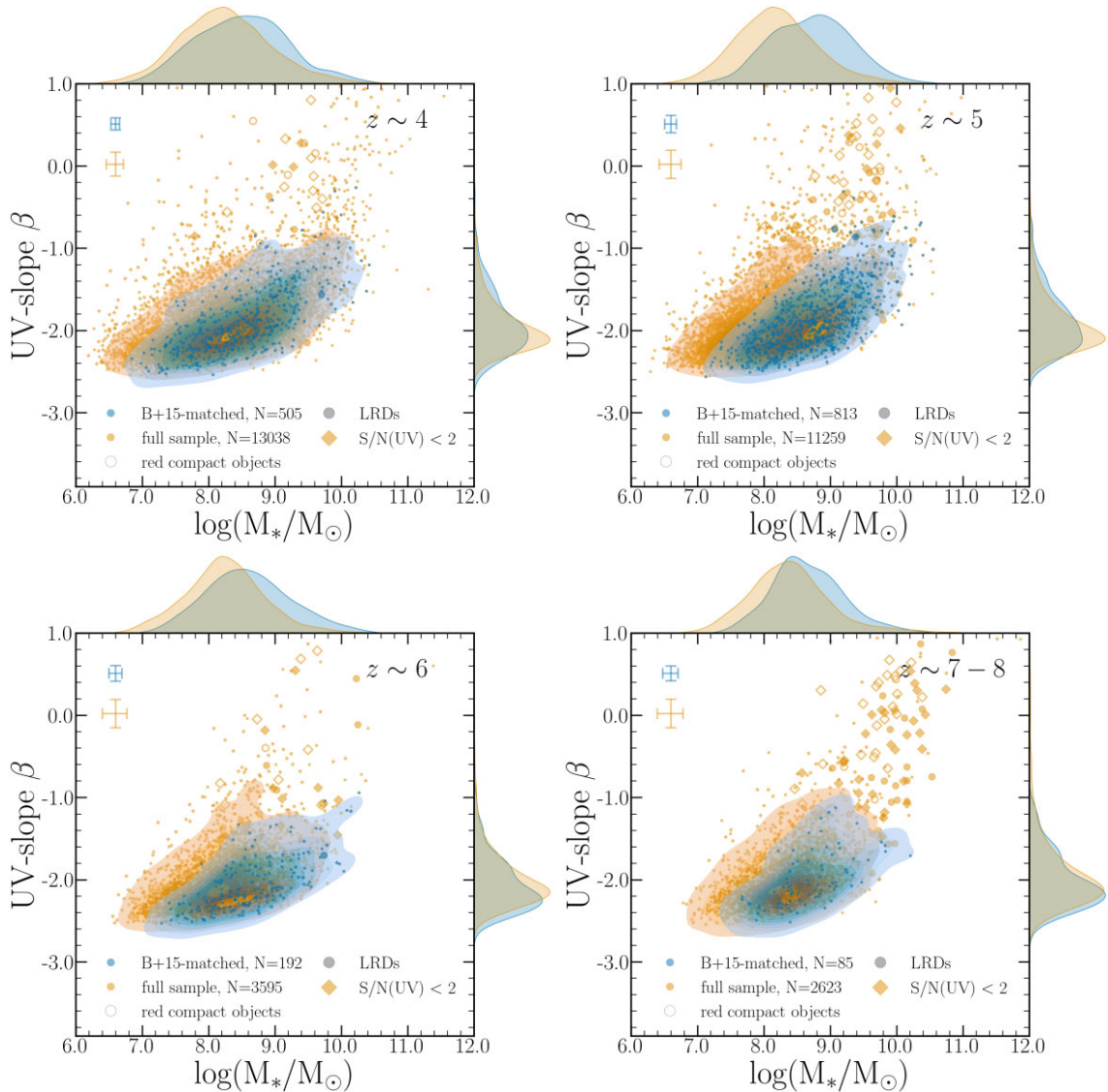

**Figure 4.** Distributions of log($M_*$) versus UV-slope $\beta$ at $z \sim 4, 5, 6$, and the combined bin $z \sim 7 - 8$ ($6.5 < z < 8.5$). The orange dots, contours, and smoothed histograms represent our full sample, while the subset of sources that are matched with the B15 sample are shown in blue. Objects plotted as diamonds are not photometrically detected in the rest-frame UV (S/N(UV)< 2); their UV-slopes are therefore simply based on the best-fitting SEDs. LRDs and red compact sources as identified in Section 2.5 are shown as filled and empty big markers (diamonds or circles). Only a few of them appear in blue, because most of them are not present in the B15 sample. The number of sources in the full sample as well as the number of B15 matched sources is indicated at the bottom of each panel. While the $\beta$-distributions on the right of each panel look similar, the mass-dependence of $\beta$ has to be taken into account for a fair comparison (see Section 3.1.2). A small number of UV-red sources ($\beta > -1.2$), not present in the B15 sample can be seen in each panel – mostly at high stellar mass, log($M_*/M_\odot$) $\gtrsim 9$ but also extending to masses as low as log($M_*/M_\odot$) $\sim 8$.

As explained in Section 2.4, $\beta$ is inferred from the best-fitting SED. We specifically highlight the objects that are not photometrically detected in the rest-frame UV (diamonds, S/N < 2 in the filter closest to 1500 Å rest frame), meaning that their $\beta$ value is based on an extrapolation of the fit in the rest-optical and not itself well constrained by the photometry. This mostly affects the reddest and most massive sources. Also, the LRDs and red compact sources (filled and empty big markers) identified in Section 2.5 are primarily located in the upper right part of the diagrams, at high masses and UV-slopes. As expected, most of them are not detected in the rest-UV and many of them were missing from the B15 sample.

The number of sources detected in this work is a factor of 10–30 × larger compared to the B15 subsample in each redshift bin. The highest increase is seen at $z \sim 4$ because only the GOODS-S field has complete *HST*/ACS *B*-band (*F435W*) coverage, allowing for the $z \sim 4$ *B*-band dropout selection in B15.

In the other three redshift bins shown, the two samples are based on roughly the same survey area over the four fields studied here. Specifically, > 80 per cent of our survey area is also covered by the HST *H* band, as can be seen in Table 1. The smoothed, normalized histograms on top of each panel show that we probe $\sim 0.25 - 0.5$ dex lower in mass compared to B15. Assuming a low-mass end slope of the SMF of $\alpha = -2$ (cf. Section 3.2.4), this results in a factor of 3 to 10 × increase in the number of galaxies. To quantify this more accurately, one would however need to take into account the relative survey depth between the underlying *JWST* and *HST* imaging for the two samples, respectively, which varies from field to field (and even within fields).

A remaining excess of galaxies detected here can be attributed to the selection functions in B15 (see their fig. 4). By design, the colour–colour LBG selections identify galaxies *around* a certain redshift, but the completeness of the selection drops as one moves further away from the central redshift of the bin. Conversely, our photo-z selection, in principle, selects galaxies uniformly across the whole redshift range in each bin.

The $\beta$-distributions, shown on the right of each panel are roughly consistent between the full sample and the B15-matched subsample. This has to be interpreted cautiously since it has been well established that $\beta$ and $M_*$ are correlated at the redshifts studied here (e.g. Finkelstein et al. 2012). As we are probing lower masses compared to B15, one might expect the $\beta$-distributions of the full sample to be slightly shifted towards lower values. However, towards the lowest masses probed, the $\beta$-values do not show a strong trend with mass anymore, but stabilize around slopes of $\beta \sim -2.2$ with a 'floor' at $\beta \sim -2.5$ induced by the models used in the SED-fitting (Section 2.4). It can therefore be inferred from Fig. 4 that there is no strong bias seen in the B15-matched sample in terms of the overall distribution of UV-slopes. Importantly though, there is a small population of UV-red galaxies seen at all redshifts which is not present in the B15-matched sample, especially beyond the contours characterizing the bulk of the sample. Those sources preferentially lie at high stellar masses, $\log(M_*/M_\odot) \gtrsim 9$, but reach down to as low as $\log(M_*/M_\odot) \sim 8$.

### 3.1.2 Fraction of UV-red sources

In order to investigate the abundance of UV-red galaxies at fixed stellar mass, i.e. to eliminate the mass-dependence of the UV-slope $\beta$, we split our sample into 'UV-blue' and 'UV-red' galaxies adopting a simple cut of $\beta > -1.2$ for the 'UV-red' sample. This roughly corresponds to a dust extinction of $A_V \sim 1$ mag. As discussed in, e.g. Wilkins et al. (2011), the value of $\beta$ depends on various physical properties of a galaxy like the metallicity, the SFH and the IMF. It is however mainly driven by dust extinction and can be taken as a good proxy for $E(B - V)$. Here, we simply want to make an observational distinction between UV-red and UV-blue galaxies which is defined somewhat arbitrarily, and is only used to illustrate differences and trends in the galaxy population. Subsequently, 'UV-red' refers to $\beta > -1.2$ and 'UV-blue' to $\beta < -1.2$ unless otherwise stated.

In Fig. 5, we show the fraction of UV-red galaxies measured in logarithmic bins of mass with bin edges defined as $\log(M_*/M_\odot) = [8, 8.5, 9, 9.5, 10.5, 11.5]$ and in four different redshift bins with bin edges $z = [3.5, 4.5, 5.5, 6.5, 8.5]$, i.e. $z \sim 4, 5, 6$ and $z \sim 7 - 8$. The wider bins at higher masses and redshifts are chosen to increase the number statistics.

We do not show the fractions in some mass and redshift bins due to the very small number of objects. In particular, at $z \sim 4$, there are 53 sources in the highest mass bin with a UV-red fraction of $\sim 95$ per cent. None of those 53 sources shows up in the B15 sample because, as noted in Section 2.6, it only covers the GOODS-S field at $z \sim 4$ while 45/53 of the ultramassive sources at $z \sim 4$ are scattered over the three other fields studied in this work (UDS, COSMOS, and EGS).

Fig. 5 shows that the UV-red fraction – as measured from our full sample – is a strong function of mass, with close to no UV-red galaxies at $\log(M_*/M_\odot) \sim 8$ and $\gtrsim 40$ per cent UV-red galaxies at $\log(M_*/M_\odot) \sim 10$, but does not show a clear dependence on redshift. The conspicuously high UV-red fraction at $z \sim 7 - 8$ and $\log(M_*/M_\odot) \sim 10$ decreases to $\sim 48$ per cent, if LRDs and red compact sources are removed, making it more consistent with the fractions measured at that mass in lower redshift bins, where including or excluding the red compact objects does not make a significant difference.

To provide another comparison, we also show the fraction of sources analogous to *HST*-dark galaxies. We use the selection cuts from Williams et al. (2024) to identify such sources requiring a red colour F150W − F444W > 2.2, and a faint magnitude at $\sim 1.5\mu$m, mag(F150W) > 27 AB. The latter cut implies that a corresponding source would typically remain undetected in CANDELS, where the *H* band at $\sim 1.6\mu$m is the reddest available filter, reaching depths $\lesssim 27$ AB (see B15). Since this selection is not actually based on *HST*-data, we subsequently refer to the corresponding sources as '*HST*-dark' in quotation marks. We show the fraction of '*HST*-dark' sources in dark red, with uncertainties in analogy to the other fractions shown.

In the highest mass-bin at $z \sim 4$, the '*HST*-dark' fraction reaches 34 per cent, indicating that at least about a third of those galaxies would be expected to be missing from *HST*-based LBG-samples. Gottumukkala et al. (2024) applied a similar selection, requiring F150W − F444W > 2.1, combined with a weaker cut in F150W, mag(F150W) > 25 AB. They show that ∼60 per cent of their selected objects are massive, dusty galaxies at $z > 3$. If we keep our red cut at 2.2 but apply their weaker magnitude cut in F150W, the fraction of such sources in the highest mass bin at $z \sim 4$ is 91 per cent, indicating that dusty, optically dark or faint galaxies completely dominate in that regime.

At higher redshifts, the number of galaxies with $\log(M_*/M_\odot) > 10.5$ drops significantly. We find eight in total at $z \sim 5$ with a UV-red fraction of 7/8 (87.5 per cent) and one-eighth sources being recovered in B15 (albeit one of the UV-red sources). None of those eight objects is selected as '*HST*-dark', but five-eighths pass the selection applied in Gottumukkala et al. (2024). In the highest two redshift bins, we do not show the fractions in the highest mass bin. At $z \sim 6$ there are only two such sources. One of them is '*HST*-dark', and they are both not present in B15. We discuss the 13 sources discovered at $z \sim 7 - 8$ with $\log(M_*/M_\odot) > 10.5$ (none of which is matched with the B15 sample) in Section 4.4 and argue that most of them likely have wrongly inferred masses and/or redshifts. Note also, that towards higher redshifts, an increasing fraction of the '*HST*-dark' sources consists of LRDs and red compact sources. In particular, the total fraction of '*HST*-dark' sources at $M_* \sim 10^{10}$ M$_\odot$ drops to 4 per cent (7 per cent) at $z \sim 6$ ($z \sim 7 - 8$) if LRDs and red compact sources are removed.

Fig. 5 further shows that the UV-red fraction is consistent between B15 and the full sample up to $\log(M_*/M_\odot) \sim 10$ and $z \sim 6$. At higher masses and/or redshifts, the fraction of red, but also of '*HST*-dark' galaxies indicate that typical *HST*-based LBG-samples may be missing a significant fraction of them. At $z \sim 7 - 8$, the UV-red fraction is close to zero in the B15 subsample at all masses while the full sample still shows an increasing UV-red fraction with mass, consistent with the trend seen at lower redshift. Interestingly, the B15 subsample also appears to be missing some UV-red galaxies at $z \sim 6$ and with $M_* \sim 10^9$ M$_\odot$.

### 3.2 Stellar mass functions

The observed abundance of UV-red or even '*HST*-dark' galaxies, in particular at the highest stellar masses, shown in the previous Section raises the question of the impact of those sources on the SMFs and in particular their high-mass end.

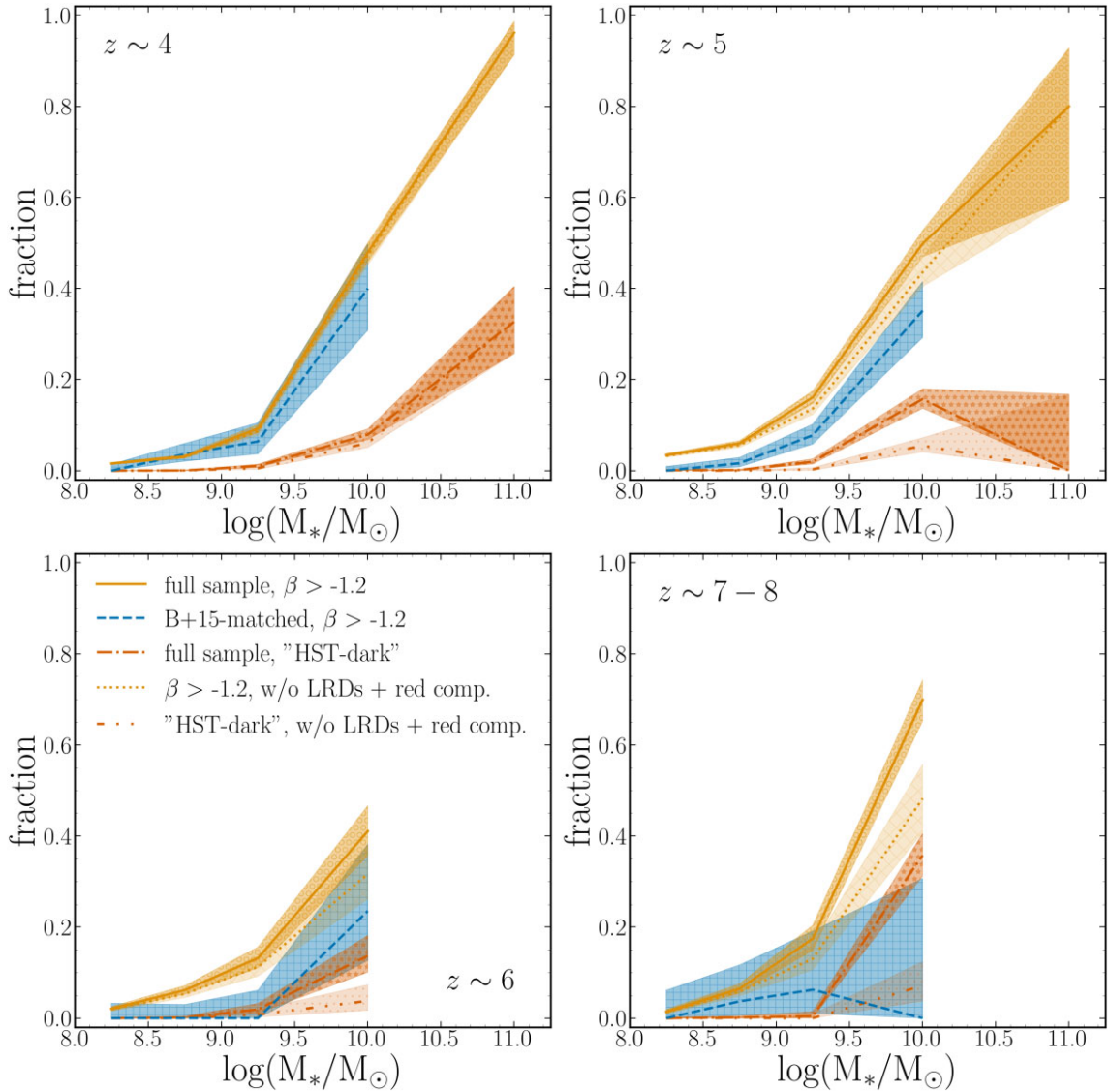

**Figure 5.** Fraction of UV-red sources, defined as sources with UV-slope $\beta > -1.2$, as a function of $\log(M_*/M_\odot)$ in four different redshift bins centred at $z \sim 4$, 5, 6 and the combined bin $z \sim 7 - 8$ ($6.5 < z < 8.5$). The orange lines represent our full sample, indicating a strong mass-dependence of the UV-red fraction in each redshift bin but no clear redshift-dependence. The lighter orange dotted lines are estimated based on a sample without the LRDs and red compact sources identified in Section 2.5. The blue lines represent the UV-red fraction for the B15 subsample. Finally, the dark red lines show the fraction of '*HST*-dark' galaxies, defined as in Williams et al. (2024), with and without LRDs and red compact sources (dash–dotted and dash–dot–dotted lines, respectively, see Section 2.5).

### 3.2.1 Binned SMFs

Fig. 6 shows the inferred SMFs in our six different redshift bins. At each redshift, the number density of galaxies is computed in mass-bins with a width of 0.5 dex, starting with the lowest mass bin in which the completeness of our sample is estimated to be > 80 per cent (see Section 2.7.2), and going up to $\log(M_*/M_\odot) = 12$. The error bars represent the 16th and 84th percentile of the 1000 SMF realizations sampled from the bagpipes posterior distributions, with Poisson uncertainty and cosmic variance added in quadrature (see Section 2.7.4). If the lower uncertainty reaches to a value $\leq 0$, we plot a downward pointing arrow instead of an error bar below the SMF point, but keep the upper error bar. In bins where we count zero galaxies, we plot an upper limit given by the Poisson single-sided $1\sigma$ upper limit from table 1 in Gehrels (1986). We also show the SMFs inferred when including LRDs (Section 2.5). Our best-fitting Schechter functions, convolved with the uncertainty in $\log(M_*)$ to account for the Eddington bias, are overplotted as the green solid lines.

Schechter fits from the pre-*JWST* literature, convolved with our inferred uncertainty distributions for consistency, are overplotted in various colours (Duncan et al. 2014; Grazian et al. 2015; Song et al. 2016; Davidzon et al. 2017; Stefanon et al. 2017, 2021; Bhatawdekar et al. 2019; Kikuchihara et al. 2020; Furtak et al. 2021; Weaver et al. 2023); the SMF from Navarro-Carrera et al. (2024) is based on *HST* data from the CANDELS catalogues, combined with some of the early public *JWST* imaging over the UDS and GOODS-S fields. The results from Wang et al. (2024a) are derived from PRIMER data, including MIRI, and those from Harvey et al. (2024) are based on a compilation of various deep fields observed by JWST (PEARLS, CEERS, GLASS, JADES GOODS-S, NGDEEP, and SMACS0723). For the latter two, we also plot markers with error bars in Fig. 6 for a more direct comparison. All the Schechter functions in Fig. 6 are only shown above their respective mass completeness limit, or in the regime where they are specified in the respective papers. Our SMFs are broadly consistent with the literature, with a few noteworthy exceptions which we list here and discuss in more detail below.

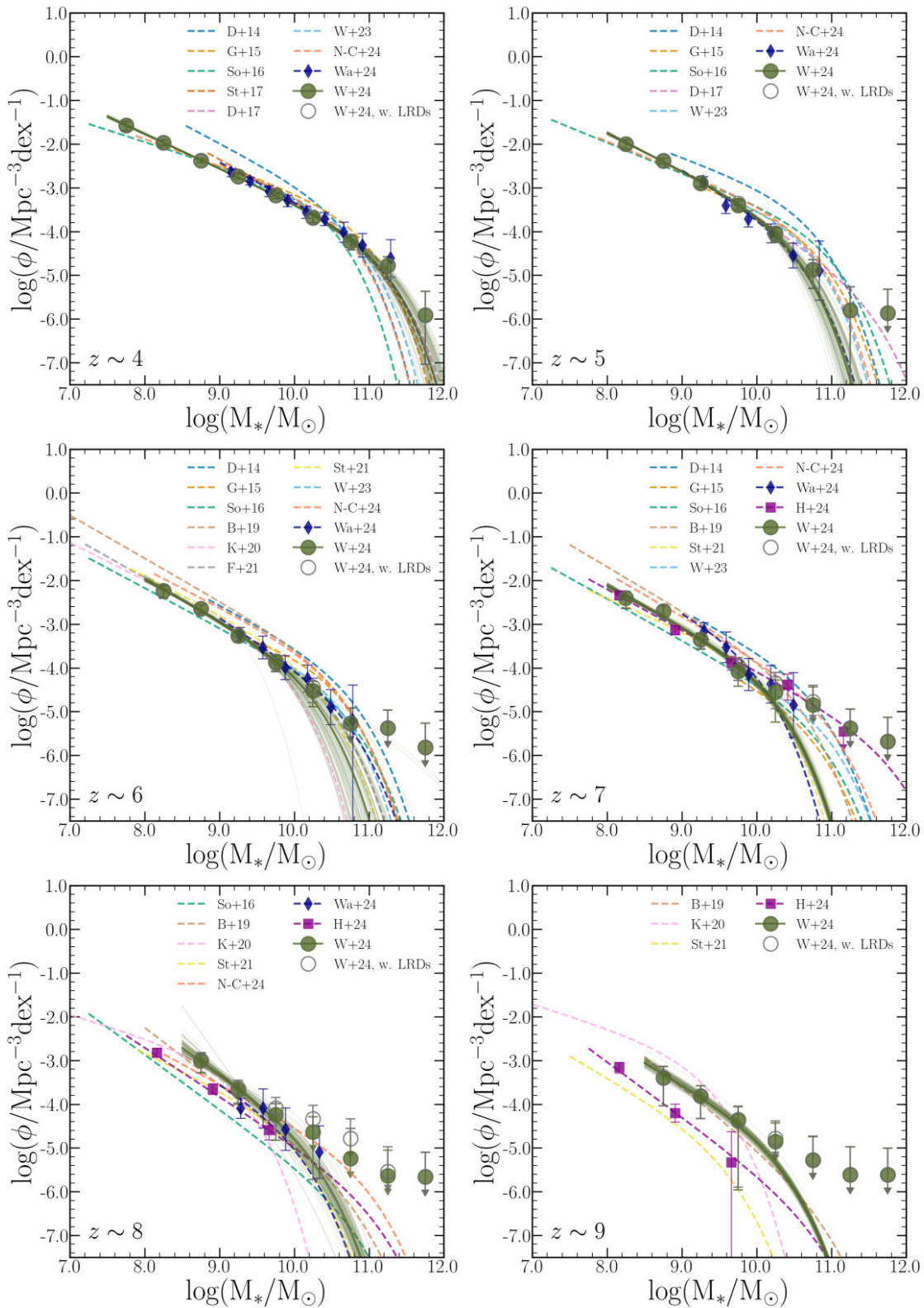

**Figure 6.** Our inferred SMFs in six redshift bins. The measured number density of galaxies in uniform mass-bins with a width of 0.5 dex is shown as the green dots. The empty grey circles represent the number densities inferred from our sample including LRDs (see Section 2.5). Our best-fitting Schechter functions are overplotted as the green solid lines, with the thin lines representing fits to 100 different realizations of our SMFs from the bagpipes posterior distributions. The dashed lines in various colours represent Schechter fits to SMFs from the literature, with recent JWST-results being additionally shown as markers. The abbreviations in the legend stand for Duncan et al. (2014) (D + 14), Grazian et al. (2015) (G + 15), Song et al. (2016) (So + 16), Stefanon et al. (2017) (St + 17), Davidzon et al. (2017) (D + 17), Bhatawdekar et al. (2019) (B + 19), Kikuchihara et al. (2020) (K + 20), Furtak et al. (2021) (F + 21), Stefanon et al. (2021) (St + 21), Weaver et al. (2023) (W + 23), Navarro-Carrera et al. (2024) (N-C + 23), Wang et al. (2024a) (Wa + 24), Harvey et al. (2024) (H+24), and this work (W + 24).

**Table 2.** Measured SMF-values in redshift bins $z \sim 4$, 5 and 6.

| Redshift bin | $\log(M_*/M_\odot)$ | $\log(\Phi/\mathrm{Mpc}^{-3}\,\mathrm{dex}^{-1})$ |
|---|---|---|
| $3.5 < z < 4.5$ | 7.75 | $-1.57^{+0.10}_{-0.12}$ |
| | 8.25 | $-1.97^{+0.06}_{-0.06}$ |
| | 8.75 | $-2.38^{+0.04}_{-0.05}$ |
| | 9.25 | $-2.74^{+0.06}_{-0.06}$ |
| | 9.75 | $-3.17^{+0.07}_{-0.08}$ |
| | 10.25 | $-3.68^{+0.09}_{-0.11}$ |
| | 10.75 | $-4.23^{+0.14}_{-0.19}$ |
| | 11.25 | $-4.78^{+0.19}_{-0.31}$ |
| | 11.75 | $-5.91^{+0.54}_{-1.13}$ |
| $4.5 < z < 5.5$ | 8.25 | $-2.00^{+0.09}_{-0.12}$ |
| | 8.75 | $-2.38^{+0.06}_{-0.07}$ |
| | 9.25 | $-2.89^{+0.08}_{-0.10}$ |
| | 9.75 | $-3.35^{+0.10}_{-0.13}$ |
| | 10.25 | $-4.04^{+0.14}_{-0.19}$ |
| | 10.75 | $-4.87^{+0.23}_{-0.43}$ |
| | 11.25 | $-5.80^{+0.54}_{-2.55}$ |
| | 11.75 | $-5.87^{+0.55}_{-\infty}$ |
| $5.5 < z < 6.5$ | 8.25 | $-2.24^{+0.12}_{-0.17}$ |
| | 8.75 | $-2.65^{+0.09}_{-0.11}$ |
| | 9.25 | $-3.26^{+0.11}_{-0.15}$ |
| | 9.75 | $-3.85^{+0.15}_{-0.21}$ |
| | 10.25 | $-4.44^{+0.20}_{-0.35}$ |
| | 10.75 | $-5.26^{+0.35}_{-\infty}$ |
| | 11.25 | $-5.38^{+0.42}_{-\infty}$ |
| | 11.75 | $-5.82^{+0.56}_{-\infty}$ |

**Table 3.** Measured SMF-values in redshift bins $z \sim 7$, 8, and 9.

| Redshift bin | $\log(M_*/M_\odot)$ | $\log(\Phi/\mathrm{Mpc}^{-3}\,\mathrm{dex}^{-1})$ |
|---|---|---|
| $6.5 < z < 7.5$ | 8.25 | $-2.40^{+0.15}_{-0.24}$ |
| | 8.75 | $-2.70^{+0.14}_{-0.20}$ |
| | 9.25 | $-3.35^{+0.14}_{-0.21}$ |
| | 9.75 | $-3.96^{+0.19}_{-0.33}$ |
| | 10.25 | $-4.35^{+0.25}_{-0.58}$ |
| | 10.75 | $-4.78^{+0.38}_{-\infty}$ |
| | 11.25 | $-5.38^{+0.43}_{-\infty}$ |
| | 11.75 | $-5.69^{+0.55}_{-\infty}$ |
| $7.5 < z < 8.5$ | 8.75 | $-3.00^{+0.18}_{-0.28}$ |
| | 9.25 | $-3.64^{+0.19}_{-0.33}$ |
| | 9.75 | $-4.09^{+0.24}_{-0.55}$ |
| | 10.25 | $-4.33^{+0.30}_{-1.39}$ |
| | 10.75 | $-4.78^{+0.45}_{-\infty}$ |
| | 11.25 | $-5.54^{+0.57}_{-\infty}$ |
| | 11.75 | $-5.66^{+0.56}_{-\infty}$ |
| $8.5 < z < 9.5$ | 8.75 | $-3.39^{+0.25}_{-0.64}$ |
| | 9.25 | $-3.81^{+0.24}_{-0.52}$ |
| | 9.75 | $-4.35^{+0.31}_{-1.54}$ |
| | 10.25 | $-4.79^{+0.40}_{-\infty}$ |
| | 10.75 | $-5.27^{+0.54}_{-\infty}$ |
| | 11.25 | $-5.61^{+0.64}_{-\infty}$ |
| | 11.75 | $-5.61^{+0.61}_{-\infty}$ |

First, at $z \sim 4$, we measure a relatively high number density of galaxies at the high-mass end compared to Schechter fits for most SMFs from the literature, but are still consistent with measurements from a number of works (e.g. Caputi et al. 2015; Weaver et al. 2023; Gottumukkala et al. 2024; Wang et al. 2024a) (see Section 4.2). This relatively high abundance of massive galaxies is not affected by LRDs or red compact objects as defined in Section 2.5, and it is not seen at $z \sim 5-6$, implying a strong evolution of the SMF high-mass end in that redshift range. The main driver of this evolution are the 53 predominantly very red galaxies at $z \sim 4$ with masses $\log(M_*/M_\odot) > 10.5$ which we have discussed separately in Section 3.1.2.

We note here that in order to compare the high-mass end of our SMFs to the literature, it is not sufficient to compare fitted Schechter functions, since it is common to fix or constrain the parameter $M^*$ at high redshifts (e.g. Davidzon et al. 2017). Additionally, due to the limited survey area, the high-mass end of the SMF is often not well constrained. The plotted downturn of the Schechter curves from the literature may therefore not accurately represent the measured number density of galaxies. We provide a more detailed comparison to the literature in Section 4.2.

Second, we infer steep low-mass end slopes $\alpha$ compared to the literature and a weak trend of steepening $\alpha$ from $z \sim 4$ to $z \sim 6$, which we discuss further in Section 3.2.4.

Third, we see an excess of galaxies at the high-mass end at $z \sim 7$ and $z \sim 8$. Due to the much smaller number statistics, our results are however still consistent with literature values, as our measurements are only upper limits. The excess is sensitive to the inclusion of LRDs (empty circles in Fig. 6) and red compact sources which contribute most significantly in this regime. We discuss the supposed extremely massive galaxies at $z \sim 7$ and $z \sim 8$ in more detail in Section 4.4 where we critically examine their implied masses and redshifts.

Fourth, our measured SMF points at $z \sim 9$ are somewhat above literature measurements by Stefanon et al. (2021) and Harvey et al. (2024), but consistent with Bhatawdekar et al. (2019) and Kikuchihara et al. (2020). Since for $\log(M_*/M_\odot) > 9.5$, the inferred number densities are largely only upper limits, our inferred Schechter fit has to be interpreted cautiously. The robustness of our $z \sim 9$ SMF is discussed separately in Section 4.5. We list all our measured SMF values in Tables 2 ($z \sim 4-6$) and 3 ($z \sim 7-9$).

### 3.2.2 Schechter function fits

When deriving our Schechter fits, we leave all three fitting parameters $\alpha$, $\Phi^*$, and $M^*$ free in the lower three redshift bins, $z \sim 4$, 5, and 6. At $z \sim 7$, 8, and 9, we fix $\log(M^*) = 10.0$ which approximately corresponds to the value fitted at $z \sim 6$. Further, we fix $\alpha = -2.0$ at $z \sim 9$, and only fit for the normalization $\Phi^*$.

As we will explore in more detail in Section 4.4, from visually inspecting SEDs and stamps, we infer that none of the sources with $\log(M_*/M_\odot) > 10$ at $z \sim 7-9$ has a robustly determined redshift and stellar mass, except for one object which is shown in Fig. B2 (bottom panel). As can be seen from the empty circles in Fig. 6, as well as in Fig. 3, a significant fraction of these are LRDs or red compact sources (Section 2.5). We therefore exclude the corresponding mass bins from our Schechter fits.

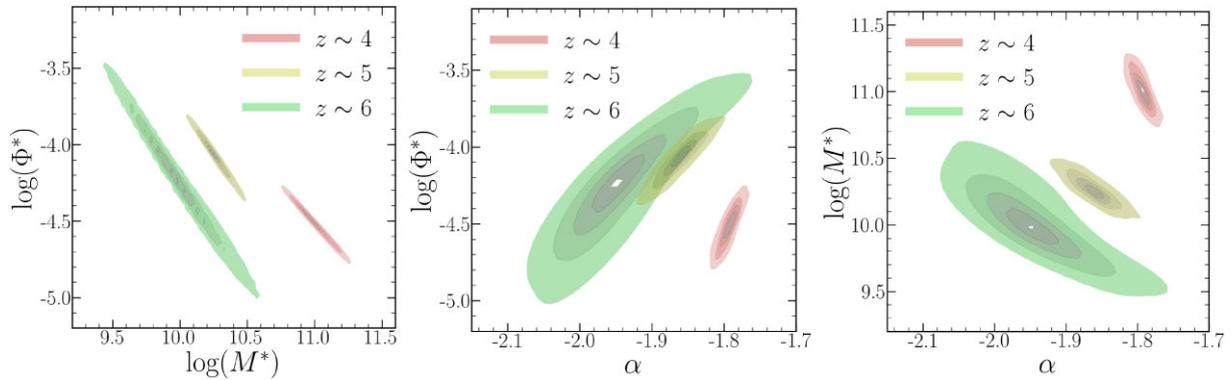

**Figure 7.** Contours of the Schechter fitting parameters, obtained from fitting to 1000 realizations of our SMFs, sampled from the bagpipes posterior distributions. The different shades represent $0.5\sigma$, $1\sigma$, $1.5\sigma$, $2\sigma$, and $3\sigma$ contours, respectively, highlighting the strong degeneracy between the three fitting parameters.

**Table 4.** Schechter fitting parameters of our inferred SMFs.

| z | $\alpha$ | $\log(\Phi^*/\mathrm{Mpc}^{-3}\,\mathrm{dex}^{-1})$ | $\log(M^*/\mathrm{M}_\odot)$ |
|---|---|---|---|
| 4 | $-1.79^{+0.01}_{-0.01}$ | $-4.52^{+0.13}_{-0.14}$ | $11.01^{+0.14}_{-0.14}$ |
| 5 | $-1.86^{+0.03}_{-0.03}$ | $-4.07^{+0.13}_{-0.14}$ | $10.26^{+0.11}_{-0.14}$ |
| 6 | $-1.95^{+0.08}_{-0.06}$ | $-4.26^{+0.36}_{-0.36}$ | $10.01^{+0.28}_{-0.36}$ |
| 7 | $-1.93^{+0.04}_{-0.04}$ | $-4.36^{+0.06}_{-0.05}$ | 10.0 (fixed) |
| 8 | $-2.16^{+0.17}_{-0.21}$ | $-4.86^{+0.19}_{-0.21}$ | 10.0 (fixed) |
| 9 | $-2.0$ (fixed) | $-4.93^{+0.08}_{-0.07}$ | 10.0 (fixed) |

In Fig. 7, we show the contours of the obtained Schechter fitting parameters at $z \sim 4-6$. This highlights the strong degeneracy between all three fitting parameters, and the rapidly increasing uncertainty on the fitting results towards higher redshifts, forcing us to fix 1 or more parameters to obtain reasonable fits at $z \sim 7-9$.

Despite this degeneracy, Fig. 7 shows a trend of a slightly steepening slope $\alpha$, as well as of an increasing $M^*$ at $z \sim 4-6$. The latter evolution is contrasted by a decreasing $\Phi^*$ from $z \sim 4$ to $z \sim 5$.

Experimenting with different constraints on $M^*$, i.e. leaving it free versus fixing it to various values in all bins, confirms that while the inferred $\Phi^*$ is very sensitive to the assumed $M^*$, $\alpha$ is quite robust to those changes at $z \sim 4-6$.

Our inferred best-fitting Schechter parameters (i.e. the median values from the contours shown in Fig. 7) are specified in Table 4.

### 3.2.3 UV-red and UV-blue mass functions

To better illustrate the contribution of UV-red ($\beta > -1.2$) galaxies discussed in Section 3.1.2 to the SMF at different redshifts, we plot SMFs for UV-red and UV-blue galaxies separately, as well as for the total sample in Fig. 8 at $z \sim 4$, 5, and 6. At $z \gtrsim 7$, the number of UV-red galaxies in our sample is not sufficient to robustly constrain the corresponding SMF. As expected from the strongly increasing fraction of UV-red galaxies with stellar mass (Fig. 5), Fig. 8 clearly shows how the population of UV-red galaxies transitions from contributing negligibly to the SMF at $\log(M_*/\mathrm{M}_\odot) \sim 9$ to the dominant population at $\log(M_*/\mathrm{M}_\odot) \sim 10$ at all redshifts. This highlights again that the observed high abundance of galaxies at the high-mass end at $z \sim 4$ is entirely driven by UV-red galaxies.

Note that we only show the UV-red SMF above a mass where our UV-red galaxy sample is estimated to be > 80 per cent complete. This threshold mass is inferred for the UV-red galaxies in analogy to the 80 per cent completeness limit for the full sample (Section 2.7.2). Therefore, we argue the observed flattening or even downturn of the UV-red SMF towards lower masses that can be seen in Fig. 8 to be real and not related to completeness effects.

### 3.2.4 Redshift evolution of the SMF

The evolution of the SMF with redshift tells us about the growth of the galaxy population with cosmic time. In Fig. 9, we show the Schechter fits to our SMFs at different redshifts in one plot. The left panel shows the fits to our full sample at $z \sim 4, 5, 6, 7, 8$, and 9 and the right panel shows Schechter fits split by UV-red and -blue at $z \sim 4$, 5, and 6. The uncertainties are again derived from the Schechter fits to the 1000 realizations of our SMFs sampled from the bagpipes posterior distributions, but we additionally perturb each measurement of $\Phi^*$ with a Gaussian random contribution from cosmic variance to represent its effect on the uncertainty of the derived SMFs.

It can be seen that the high-mass end changes significantly from $z \sim 4$ to $z \sim 5$ which is reflected in the inferred values of $\log(M^*/\mathrm{M}_\odot) = 11.01^{+0.14}_{-0.14}$ at $z \sim 4$ and $\log(M^*/\mathrm{M}_\odot) = 10.26^{+0.11}_{-0.14}$ at $z \sim 5$. When interpreting this shift in $M^*$, the degeneracy between $M^*$ and $\Phi^*$ however has to be taken into account, which in this case leads $\Phi^*$ to *increase* from $z \sim 4$ to $z \sim 5$ (see Table 4). The right panel illustrates that while the SMF of the UV-blue galaxies only shows a modest evolution from $z \sim 4-6$, the number density of UV-red massive galaxies evolves strongly in this redshift range. We further discuss this rapid evolution in Section 4.2.

While we cannot simultaneously constrain *both* $\Phi^*$ and $M^*$ at $z \gtrsim 4$ (see Fig. 7, we can better constrain the redshift evolution of the low-mass end slope $\alpha$ and show this in Fig. 10, comparing to various results from the literature. For our sample, the uncertainty is artificially reduced at $z \sim 7, 8$ where we fix the value of $M^*$ in the fitting, and we do not show the results at $z \sim 9$ where we also fix $\alpha$.

We note that our inferred values of $\alpha$ are relatively steep overall, reaching as low as $\alpha \sim -2$ at $z \sim 6$. Compared to those shown from the literature, only Davidzon et al. (2017) infer even lower values at $z \sim 4-5$.

The weak trend of steepening slopes that can be seen from $z \sim 4$ to $z \sim 6$ is consistent with the trend seen in Song et al. (2016) and the recent work by Navarro-Carrera et al. (2024), who however infer somewhat higher values of $\alpha$ in this redshift range. Our measured values of $\alpha$ at $z \sim 7-8$ are consistent with various literature results within the typically large uncertainties. They are however affected by the fixed $M^*$ in these redshift bins due to the correlation between $\alpha$ and $M^*$ (see Fig. 7).

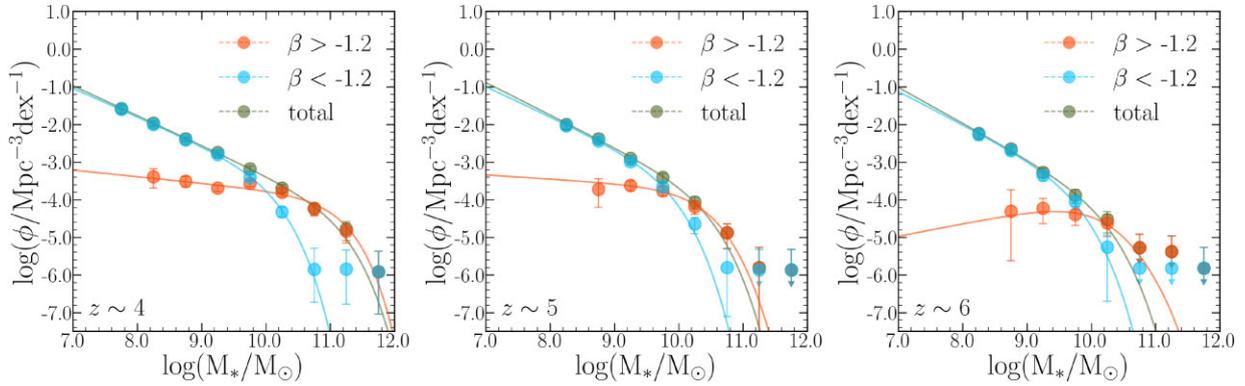

**Figure 8.** SMFs for UV-red ($\beta > -1.2$) and UV-blue ($\beta < -1.2$) galaxies in three redshift bins, respectively. The overplotted dashed lines represent fitted Schechter functions, derived in analogy to the fits described in Section 3.2.1, but fixing $M^*$ to the value inferred from the full SMF when fitting the UV-red sources. At all displayed redshifts, UV-red galaxies dominate the SMFs at log $M_*/M_\odot > 10.5$. UV-red galaxies have a much shallower low-mass slope and are thus sub-dominant at lower masses. The remarkable high-mass end at $z \sim 4$ consists entirely of UV-red galaxies.

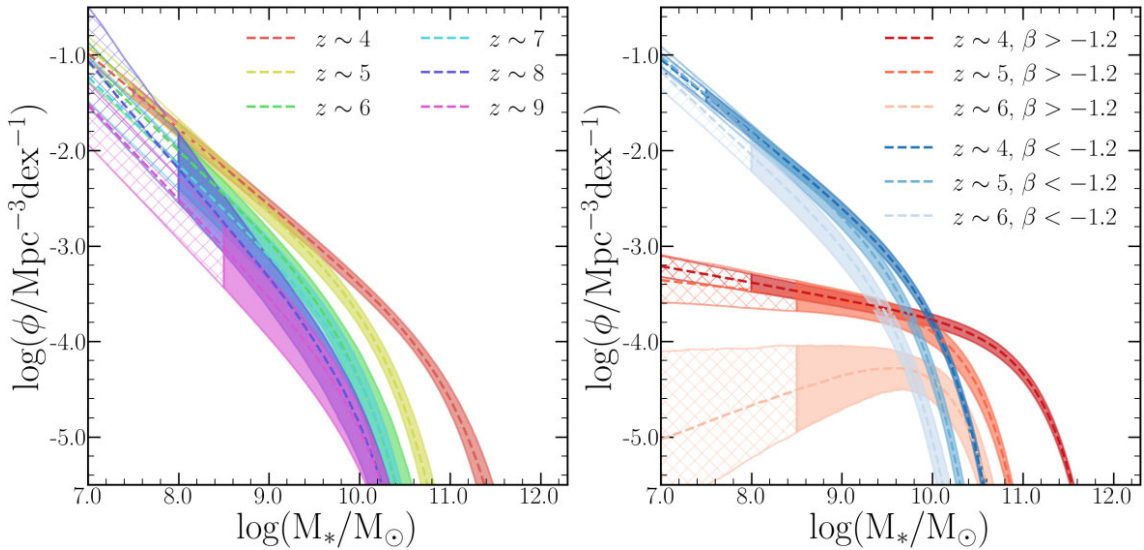

**Figure 9.** Best-fitting Schechter functions to our SMFs are shown for the full sample at $z \sim 4, 5, 6, 7, 8$, and 9 on the left and for the subsets of UV-red ($\beta > -1.2$) and UV-blue ($\beta < -1.2$) galaxies at $z \sim 4, 5$, and 6 on the right. The shaded regions indicate the 16th and 84th percentiles inferred from fitting to 1000 realizations of the SMFs with $\Phi^*$ being additionally perturbed to account for the uncertainty due to cosmic variance. The checked area on the left represents the mass range where our sample is <80 per cent complete according to Section 2.7.2 in each redshift bin respectively.

### 3.2.5 Comparison to models and simulations

We provide a comparison of our SMFs to several models and simulations in Fig. 11. First, we use the python package hmf (Murray, Power & Robotham 2013; Murray et al. 2021) to compute halo mass functions (HMFs) in each redshift bin, using the model from Tinker et al. (2008). We multiply each HMF by a constant baryon fraction of $f_b = 0.16$ (Jarosik et al. 2011) and a baryon-to-star conversion efficiency $\epsilon = 0.1, 0.3, 0.5$, and 1, respectively. Since this simple computation does not take into account any feedback effects, the resulting SMF represents a theoretical upper limit for the SMF at a given redshift and for a given baryon conversion efficiency $\epsilon$. The area corresponding to $\epsilon > 1$ is shown as the grey shaded area in each panel in Fig. 11 and is not physically allowed given the assumed cosmology and HMF model. At low masses, our SMFs remain well below the $\epsilon = 0.1$ curve due to the mentioned feedback effects. However, the high-mass end at $z \sim 5$ is consistent with $\epsilon = 0.1 - 0.3$. With increasing redshift, higher efficiencies are suggested to account for the observed abundance of galaxies at the high-mass end. This reaches to $\epsilon \sim 0.3$ or beyond at $z \sim 7 - 9$, based on the highest mass bin where we can measure the number density, rather than just provide an upper limit, and the best-fitting Schechter functions. This trend is consistent with recent findings by Chworowsky et al. (2023) who measured the number density of massive galaxies in CEERS. Somewhat against that trend, the high-mass end at $z \sim 4$ is more consistent with $\epsilon \sim 0.3$.

Also shown in the figure is a range of SMFs from various simulations, including semi-analytic models (DELPHI, Mauerhofer & Dayal 2023; Santa Cruz Yung et al. 2019), hydrodynamic simulations (FIRE-2 Ma et al. 2018; FLARES Lovell et al. 2021; THESAN Kannan et al. 2022; ASTRID Bird et al. 2022; SPHINX Katz et al. 2023), and the feedback-free starburst (FFB) model proposed by Dekel et al. (2023), with observational predictions in Li et al. (2023). For the latter, we assume a maximum star formation efficiency $\epsilon_{max} = 0.2$, where $\epsilon_{max}$ is defined as the baryon fraction times the halo growth rate divided by the mean SFR (see equation 4 in Li et al. 2023). It therefore differs from our definition of $\epsilon$ above. Overall, most of the SMFs from simulations agree well with our measurements out to at least $z \sim 6$. An exception is the SPHINX SMF which lies significantly above our SMF at $z \sim 5$ and $z \sim 6$. Note that the turnover at the low-mass end of the SPHINX SMF is due to the applied cut of SFR> $0.3 M_\odot yr^{-1}$ in the catalogue from Katz et al. (2023), which has no impact at higher masses.

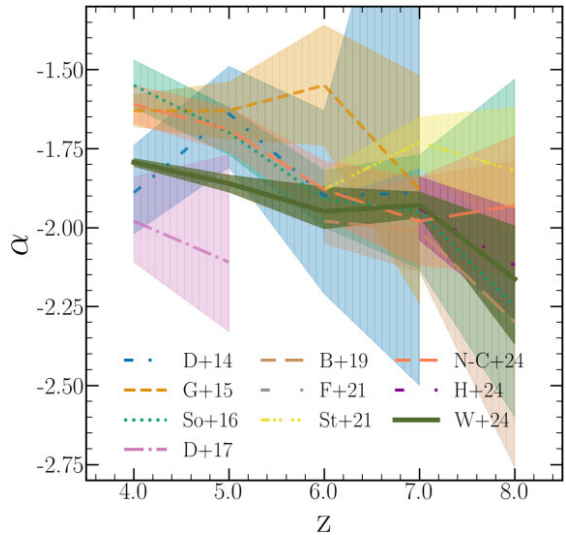

**Figure 10.** Redshift evolution of the low-mass end slope $\alpha$ inferred from our best-fitting Schechter functions. Various literature results are shown with the shaded regions representing 16th and 84th percentiles or $1\sigma$ uncertainties. The abbreviations stand for Duncan et al. (2014) (D + 14), Grazian et al. (2015) (G + 15), Song et al. (2016) (So + 16), Davidzon et al. (2017) (D + 17), Bhatawdekar et al. (2019) (B + 19), Furtak et al. (2021) (F + 21), Stefanon et al. (2021) (St + 21), Navarro-Carrera et al. (2024) (N-C+23), and this work (W + 24). The uncertainties shown for our measurement do not include cosmic variance.

Towards higher redshifts $z \gtrsim 7$, the scatter between different models and simulations increases. In particular, the SMFs from the ASTRID simulation, the Santa Cruz SAM, and THESAN lie somewhat below our measured values at the highest redshifts ($z \sim 7 - 9$ for ASTRID, and $z \sim 8 - 9$ for the other two). FLARES lies slightly below our measurements at $z \sim 9$, while DELPHI, baselined against all available dust observations at $z \sim 5 - 7$ (Dayal et al. 2022), is consistent with our measurements at all redshifts. The same is true for FIRE-2 which does however only probe the low-mass end of the SMF at high redshifts. Consequently, their SMF hardly overlaps with our measured values at $z \sim 9$. Interestingly, SPHINX becomes more consistent with our measurements towards higher redshifts and matches our observations at $z \sim 9$. The FFB model matches our observations well at all redshifts for $\epsilon_{max} = 0.2$. Compared to the other models, it predicts a higher number density of high mass galaxies, especially at $z \sim 8 - 9$. This is consistent with but cannot be confirmed by our observations (see also Section 4.5). A more detailed discussion of the reasons for and implications of the observed consistencies and differences between simulations and observations is beyond the scope of this work.

### 3.2.6 Cosmic stellar mass density

Another quantity commonly used to characterize the global evolution of the galaxy population is the cosmic stellar mass density (CSMD) $\rho_*$, defined as the integral over the SMF multiplied with $M_*$

$$\rho_*(z) = \int_{M_{min}}^{M_{max}} \Phi_z(M)\, M\, dM, \quad (7)$$

where $\Phi_z$ is the inferred SMF in the redshift bin centred at $z$ and $(M_{min}, M_{max})$ are the integration boundaries which in all the literature results to which we compare our measurements are defined as $M_{min} = 10^8 M_\odot$ and $M_{max} = 10^{13} M_\odot$.

In Fig. 12 we show our measurements of $\rho_*$ obtained from integrating our best-fitting Schechter functions over the same range in each redshift bin. Overall, our results are consistent with literature values within the uncertainties. They show a relatively rapid evolution from $z \sim 4 - 6$ which then becomes more shallow at $z \sim 7 - 9$, causing our measurements to lie at the upper edge but still consistent with previous estimates at $z \sim 8 - 9$. Specifically, at $z \sim 9$ our measurements lie $0.6 - 0.8$ dex higher than what was inferred in Stefanon et al. (2021) and recently in Harvey et al. (2024). We again caution against overinterpreting the SMF and therefore the CSMD at $z \sim 9$, and refer the reader to Section 4.5 for more details. Further, we discuss possible sources of systematic uncertainty that may be more important at the highest redshifts probed, such as an evolving IMF or bursty SFHs, in Section 4.3.

Fig. 12 also shows the CSMD split into UV-red and -blue galaxies at $z \sim 4, 5$, and 6. Consistent with the insights from Fig. 9, the contribution of UV-red galaxies to $\rho_*$ evolves strongly from $z \sim 4$ to 6. UV-red galaxies with dust extinction $A_V \gtrsim 1$ (see Section 3.1.2) *dominate* the CSMD at $z \sim 4$ for $M_* > 10^8 M_\odot$, contributing $\sim 60$ per cent. They transition to becoming a sub-dominant population at $z \sim 6$, where they only contribute $\sim 20$ per cent. This reflects a rapid build-up of dusty galaxies across this redshift range.

## 4 DISCUSSION

### 4.1 Red galaxies prior to and with *JWST*

By design, LBG samples were missing a so-far unknown fraction of red galaxies, as those are very faint or completely undetected in even the deepest available optical to NIR imaging. This includes, e.g. sub-mm galaxies (SMGs) at $z > 3$ (e.g. Dunlop et al. 2004; Chapman et al. 2005; Riechers et al. 2013; Zavala et al. 2021), or the less extreme so-called H-dropouts, *HST*-dark or *HST*-faint galaxies, optically faint galaxies (OFGs) or *HST* to IRAC Extremely Red Objects (HIEROs) (e.g. Huang et al. 2011; Wang et al. 2016, 2019; Alcalde Pampliega et al. 2019; Xiao et al. 2023).

The red colour of these sources can be explained by dust or old/quiescent stellar populations creating a strong Balmer break and lacking significant rest-frame UV emission from young stars. Since quiescent galaxies have only been observed out to $z \sim 4$ and they are expected to be extremely rare at those redshifts (e.g Carnall et al. 2020; Long et al. 2024; Valentino et al. 2023), we expect dusty star-forming galaxies to contribute more significantly to our SMFs. The contribution of $K$-band- or IRAC-selected red galaxies to the high-mass end of the SMF at $z \gtrsim 3$ was discussed in e.g. Marchesini et al. (2010), Caputi et al. (2011, 2015), and Stefanon et al. (2015), indicating a significant contribution, in particular at $z \gtrsim 4$.

*JWST* has revolutionized the field of *HST*-dark galaxies as it provides an enormous improvement in terms of spatial resolution and sensitivity compared to *Spitzer*/IRAC at $3 - 5 \mu m$ yielding much better constraints on the physical properties of red galaxies. Various authors have exploited imaging from the first year of *JWST* to investigate such galaxies, typically selected via a red colour between *F*444*W* or *F*356*W* and either the *HST H* band or *F*200*W*/*F*150*W* which probe similar wavelengths (e.g. Barrufet et al. 2023; Gómez-Guijarro et al. 2023; Nelson et al. 2023; Pérez-González et al. 2023; Rodighiero et al. 2023; Gottumukkala et al. 2024; Williams et al. 2024). While the dusty star-forming nature of the higher redshift objects ($z \gtrsim 3$) in those samples has largely been confirmed with *JWST*, accurately constraining their stellar masses, redshifts and star formation histories remains difficult from photometry alone. Follow-up spectroscopy with NIRSpec as well as constraints at longer wavelengths by either MIRI (see the recent work by Pérez-González et al. 2024) or in the sub-mm domain (see Labbe et al. 2023a) will provide further insights in the future (see also Section 4.4).

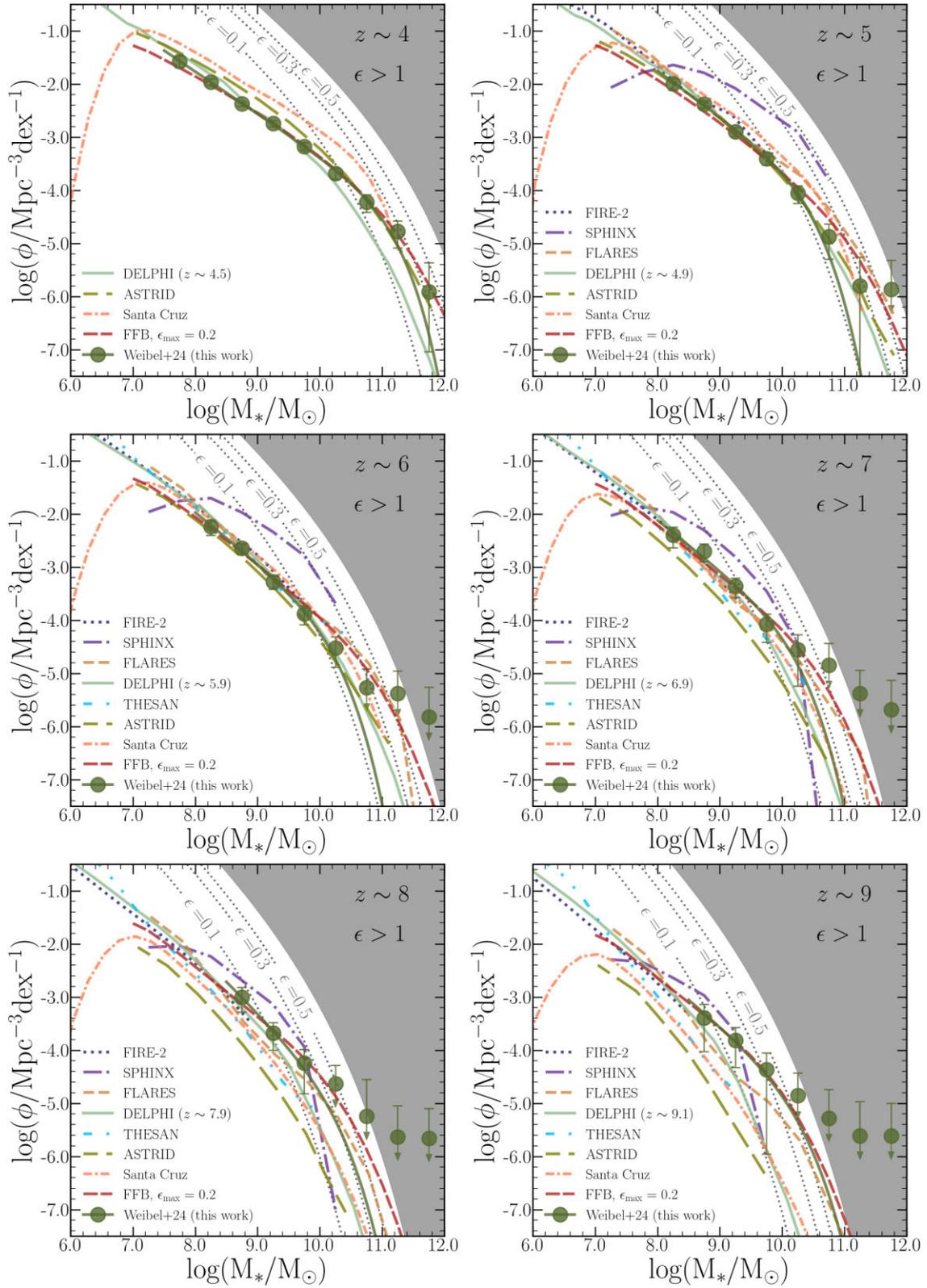

**Figure 11.** Comparison of our SMFs to various models and simulations. In each panel, we show our measured SMFs and the best-fitting Schechter functions in green. For comparison, we show theoretical upper limits on our SMFs, inferred from a Tinker et al. (2008) HMF, multiplied with a baryon fraction of $f_b = 0.16$ and various values of the baryon conversion efficiency $\epsilon = 0.1, 0.3, 0.5$, and 1. The grey shaded area corresponds to $\epsilon > 1$ and is not physically possible given the assumed cosmology and HMF model. We further overplot SMFs from various semi analytic models (DELPHI, Mauerhofer & Dayal 2023; Santa Cruz Yung et al. 2019), hydrodynamic simulations (FIRE-2 Ma et al. 2018; FLARES Lovell et al. 2021; THESAN Kannan et al. 2022; ASTRID Bird et al. 2022; SPHINX Katz et al. 2023), and the FFB model from Dekel et al. (2023); Li et al. (2023).

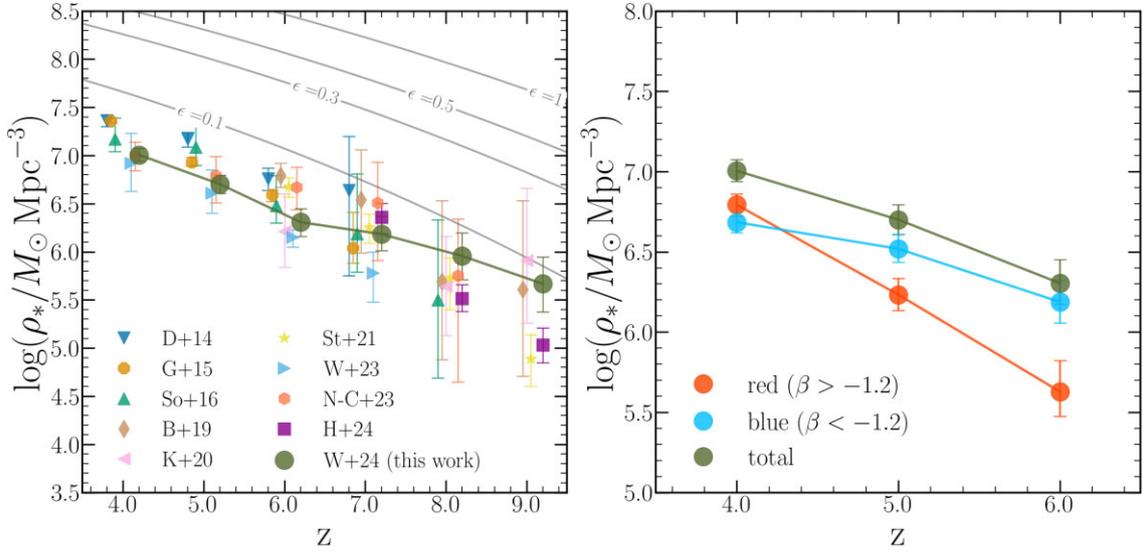

**Figure 12.** Inferred CSMD obtained from integrating our best-fitting Schechter SMFs above $M_{min} = 10^8$ M$_\odot$. On the left, various points from the literature are shown for comparison to our total CSMD. Each marker is displaced by an arbitrary offset in the x-direction for better visual separation of the different results. We show theoretical SMFs for different values of $\epsilon$, obtained from multiplying the HMF from Tinker et al. (2008) by a constant baryon fraction and $\epsilon$, and then integrating down to $M_{min} = 10^8$ M$_\odot$ according to equation (7). On the right, the CSMD is shown for our UV-red ($\beta > -1.2$) and UV-blue ($\beta < -1.2$) subsamples, showing a rapid evolution in the CSMD of UV-red galaxies, contrasted by a more shallow evolution in the CSMD of UV-blue galaxies.

To compare our sample of UV-red galaxies to pre-*JWST* selections of sub-mm and optically faint or *HST*-dark galaxies, we focus on the GOODS-S field. In Fig. 13, we plot our sample galaxies ($3.5 < z < 9.5$) in GOODS-S in a *F*150*W* − *F*444*W* versus *F*444*W* colour–magnitude diagram and colour-code each source with its inferred UV-slope $\beta$ (Section 2.4). Furthermore, we indicate sources selected as sub-mm galaxies, OFGs, *HST*-dark galaxies or red galaxies with *JWST* by various authors in the same field. It should be noted that the sample considered in this work only covers a fraction of GOODS-S defined by the JADES DR2 footprint, which is why the overlap with other samples in this field is limited.

Not surprisingly, Fig. 13 shows that all galaxies passing a typical OFG or *HST*-dark selection are also part of our UV-red sample. However, there is a much larger number of galaxies selected as UV-red in this work that would not qualify as optically faint, indicating that our UV-red sample is much broader than typical OFG or *HST*-dark samples. Below the 'typical' *F*150*W* − *F*444*W* = 2.2 selection cut (e.g. Williams et al. 2024), our UV-red galaxies can be roughly split into two subsets, one at mag(*F*444*W*)$\lesssim$ 27 and the other at mag(*F*444*W*)$\gtrsim$ 27. The former typically have detected flux in *F*150*W* and robustly inferred UV-slopes and are simply not 'red enough' in $F150W - F444W$ to qualify as OFGs or '*HST*-dark'. The latter sources often drop out of $F150W$ and their UV-slope $\beta$ is inferred from the best-fitting SED which is not well-constrained in the rest-frame UV. Therefore, their $\beta$ values have to be interpreted cautiously. We emphasize that since those sources are at the faint-end of our sample and have typical masses of around $10^8$ M$_\odot$, they do not contribute significantly to the SMFs shown above.

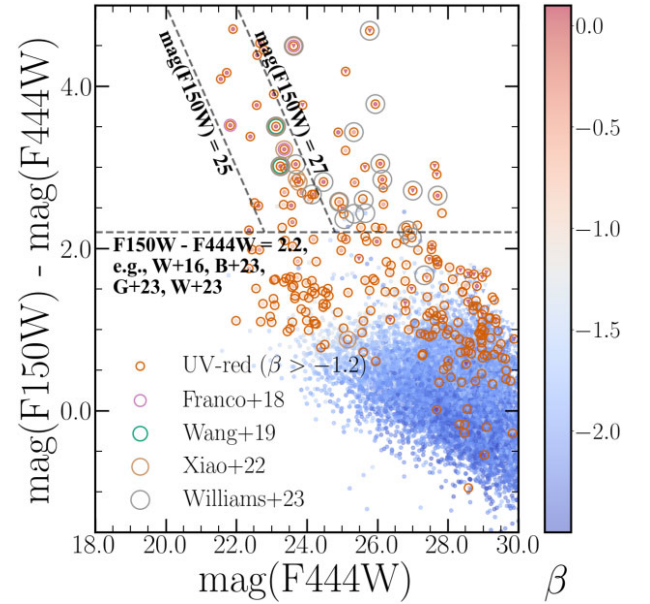

**Figure 13.** *F*150*W* − *F*444*W* versus *F*444*W* colour–magnitude diagram of our sample galaxies ($3.5 < z < 9.5$) in the GOODS-S field (JADES DR2 footprint), colour-coded with the inferred UV-slope $\beta$. Galaxies selected as UV-red ($\beta > -1.2$) are highlighted with red circles. Other circles highlight sources that have been selected as ALMA-detected sources with no *HST*-counterparts (Franco et al. (2018), violet circles), H-dropouts with ALMA detections (Wang et al. (2019), green circles), OFGs (Xiao et al. (2023), brown circles), and as *JWST*-red sources analogous to H-dropouts/*HST*-dark galaxies (Williams et al. (2024), grey circles) in the same field. Non-detections in *F*150*W* are assigned a $2\sigma$ upper limit and plotted as triangles. The grey dashed lines indicate typical selection boxes, similar to those used in e.g. Wang et al. (2016) (W + 16), Barrufet et al. (2023) (B + 23), Gottumukkala et al. (2024) (G + 23), and Williams et al. (2024) (W + 23).

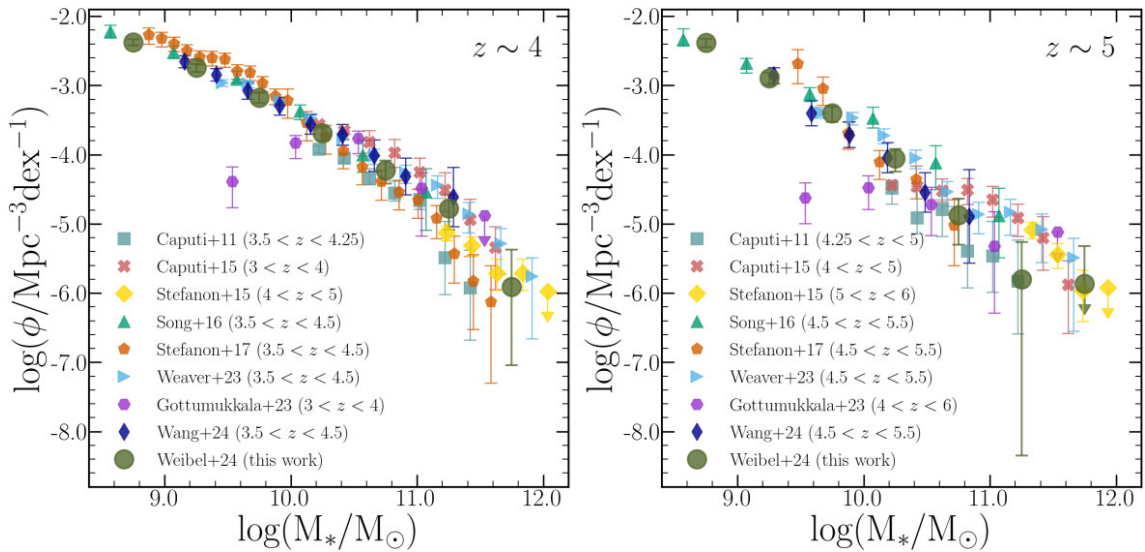

**Figure 14.** High-mass end of the inferred SMFs at $z \sim 4$ (left panel) and $z \sim 5$ (right panel), compared to various results from the literature as specified in the legends. The green markers representing this work are equivalent to those shown in Fig. 6.

We conclude that our sample includes OFG/*HST*-dark galaxies studied in the pre-*JWST* literature and complements them with additional fainter UV-red sources, as well as sources with substantially less extreme colours compared to typical OFGs/*HST*-dark galaxies. We apply a relatively mild selection cut of $\beta > -1.2$, which selects sources with $A_V \gtrsim 1$ that are dusty but not as extreme as OFGs/*HST*-dark galaxies which typically show $A_V \gtrsim 2$ (e.g. Gottumukkala et al. 2024). This shows that we reached the goal of this work to provide a complete census of galaxies at $z \sim 4 - 9$, which also includes the previous, colour-selected OFGs.

### 4.2 The high-mass end of the SMF at z ∼ 4 − 5

To explore in more detail the high-mass end of our SMFs, we present our SMFs again in Fig. 14 at $z \sim 4$ (left panel) and $z \sim 5$ (right panel), directly comparing the actual measurements at $\log(M_*/M_\odot) \gtrsim 9$ to various results from the literature. In particular, we include comparisons to Caputi et al. (2011), Caputi et al. (2015), Stefanon et al. (2015) and Stefanon et al. (2017) who constructed SMFs based on *Spitzer*/IRAC 4.5μm selected or complemented galaxy samples, rather than building on *HST*-only selected samples that might miss OFGs. Further, we overplot the SMFs from Weaver et al. (2023) who included ground-based H- and K$_S$-band imaging in their detection image, as well as the *JWST*-based SMFs presented in Gottumukkala et al. (2024) who specifically inferred the SMF of OFGs only, selected through a red $F150W - F444W$ colour in CEERS, and SMFs from Wang et al. (2024a) who constrained the high-mass end of the SMF using NIRCam + MIRI data from PRIMER. Additionally, we show results from Song et al. (2016) who provide measurements of the SMF based on a sample selected from the *HST J* and *H* bands, without adding sources selected at longer wavelengths.

While our measurements at $z \sim 4$ are near the upper edge of the range spanned by the displayed literature results at $\log(M_*/M_\odot) \gtrsim 11$, they are formally consistent with all of them within error bars out to the highest masses, meaning that the high number density at the high-mass end of the SMF at $z \sim 4$ found in this work was measurable prior to *JWST*, in particular if sources selected at wavelengths beyond the range of *HST*, either from *Spitzer*/IRAC or from ground-based *K*-band photometry, were included (e.g. Caputi et al. 2015; Weaver et al. 2023). Our SMF is also consistent with the SMF computed by Gottumukkala et al. (2024) at $\log(M_*/M_\odot) \gtrsim 10.5$, re-emphasizing the point that OFGs completely dominate the high-mass end of the SMF. We note that due to the similar data processing and photometric measurements between Gottumukkala et al. (2024) and this work, the consistency between the two may not be surprising. Remarkably though, Gottumukkala et al. (2024) fully recovered the high-mass end of the SMF based on a sample of OFGs selected from just the CEERS field.

Unlike the case at $z \sim 4$, the space density we measure at $z \sim 5$ and $M_* \sim 10^{11}$M$_\odot$ is near the low end of previous results from the literature, but with sufficiently large uncertainty, so that our measurement is still consistent with all the displayed results. This is because our SMF evolves strongly from $z \sim 4$ to $z \sim 5$ at the high-mass end and the error bars grow significantly due to increasing cosmic variance and decreasing number counts. We emphasize the consistency with Gottumukkala et al. (2024) at $z \sim 5$ who thus also find a strong evolution in the number density of OFGs from $z \sim 5$ to $z \sim 4$.

As outlined in the following Section, we further argue the inferred redshifts and stellar masses at $z \sim 4 - 6$ to be robust to various possible sources of systematic uncertainty. Therefore, the question remains as to the physical mechanisms behind this evolution.

Naively, it implies that galaxies which are already massive at $z \sim 5$ grow very efficiently from $z \sim 5$ to $z \sim 4$. This may be related to more efficient cooling, higher gas accretion rates or merger rates of massive galaxies at these redshifts. We note that it is exactly in this redshift range that we observe the first massive quiescent galaxies in the Universe (e.g. Carnall et al. 2020, 2023b; Santini et al. 2021; Valentino et al. 2023; Long et al. 2024). Adopting the rest-frame UVJ colour-cuts from Williams et al. (2009) to identify quiescent galaxies in our sample, based on the rest-frame colours estimated by bagpipes, we find 15 quiescent galaxy candidates in total. All of them are in our $z \sim 4$ bin, and seven of them have $\log(M_*/M_\odot) > 10.5$. If we slightly soften the selection criteria and retain sources whose 16th and 84th percentiles of the rest-frame colours are consistent with the selection box, those numbers increase to 33 and 10. From visual inspection (see Section 4.4), we find that precisely those 10 galaxies out of the 53 with $\log(M_*/M_\odot) > 10.5$ at $z \sim 4$ show red SEDs, elliptical morphologies and low inferred sSFRs from bagpipes, characteristic of quiescent galaxies. The strong evolution of the CSMD of UV-red galaxies shown in Fig. 12 (right panel) also points to a rapid build-up of dust in the most massive systems from $z \sim 5$ to $z \sim 4$, leading to a higher obscured fraction of UV-light and star formation.

### 4.3 Systematic uncertainties in the inferred stellar masses

While we have discussed and taken into account random measurement and SED-fitting uncertainties in our stellar mass estimates (see Section 2.7.4), there are several sources of systematic uncertainty which we wish to discuss here.

First, the choice of an SFH model may have some impact on the inferred stellar masses. For example, fitting the SEDs of simulated galaxies, Ciesla, Elbaz & Fensch (2017) found mean errors of up to 40 per cent with the delayed-$\tau$ SFH out to $z \sim 5$. At higher redshifts, this effect may however increase if the SFHs become more bursty as has been recently proposed by various authors (e.g. Ciesla et al. 2023; Cole et al. 2023; Looser et al. 2023). Investigating galaxies at $6.5 < z < 13.5$, Harvey et al. (2024) found typical offsets of $\sim 0.3 - 0.4$ dex between stellar masses inferred using a delayed-$\tau$ and a non-parametric SFH, with the latter yielding higher masses.

We reran our SED-fitting using a double power-law parametrization of the SFH with an additional window-function burst. This allows for more flexibility, and adds two more free parameters compared to our fiducial delayed-$\tau$ model. We find a median offset of $\sim 0.1$ dex, with higher masses inferred with the double power-law + burst model. This offset however reverses towards the highest masses, where for $\log(M_*/M_\odot) > 11$, the masses decrease by $\sim 0.2$ dex with the new model. Overall, the effect of choosing a different SFH model on the main results of this paper is marginal. While the delayed-$\tau$ model is certainly too simplisitic to accurately represent SFHs of individual high-redshift galaxies, it appears to provide statistically robust measurements of the SMF. More work is required to examine in detail the variety of SFHs at high redshifts and how accurately they can be constrained based on photometry and SED-fitting.

For example, as has been shown by the work of Giménez-Arteaga et al. (2023, 2024), spatially resolved SED-fitting of lensed galaxies at $z \sim 5 - 9$ increases their inferred stellar masses by as much as 0.5–1 dex compared to integrated aperture photometry, because the young stellar populations that may have formed in recent bursts outshine the older stellar populations in the integrated photometry. This analysis has only been performed for a handful of sources which may not be representative of the galaxy population at any given redshift and accounting for this effect is not possible within a standard SED-fitting framework.

Another possibility discussed in the literature is that the IMF changes as a function of the metallicity or the dust temperature (Chon, Omukai & Schneider 2021; Sneppen et al. 2022). Both mechanisms would imply that the IMF becomes more top-heavy towards higher redshifts. A more top-heavy IMF decreases the mass-to-light ratio and would lead us to overestimate stellar masses. If the IMF changes at $z \sim 8 - 9$, this may account for the measured slow evolution of the SMF and the CSMD in this redshift range. Since we cannot constrain the IMF based on the available data, we do not further explore this possibility.

In particular for the reddest, most dusty and massive systems, having access to the rest-frame near-infrared emission which can be probed by MIRI may significantly improve constraints on the stellar mass. Williams et al. (2024) performed a detailed analysis of extremely red galaxies in GOODS-S, including various NIRCam medium-band filters and MIRI data in seven filters, concluding that with MIRI the inferred stellar masses show a median decrease of $\sim 1$ dex, compared to using HST + NIRCam alone for sources initially showing $M_* > 10^{10} \, M_\odot$. Looking at a broader sample of galaxies at $4 < z < 9$ in CEERS, Papovich et al. (2023) find that with MIRI, the inferred stellar masses decrease by 0.25 (0.37) dex at $4 < z < 6$ ($6 < z < 9$). Using MIRI-data from the PRIMER survey, Wang et al. (2024a) specifically investigated the impact of including MIRI on the stellar masses of massive galaxies ($M_* > 10^{10} \, M_\odot$), and found no significant impact at $z \sim 4$, a median decrease of the MIRI-inferred masses by $0.1 - 0.2$ dex at $z \sim 5 - 6$, and a more significant decrease of up to 0.5 dex at $z \sim 7 - 9$.

For our sample, this means that in particular some of the extreme masses inferred for the very red sources at $z \sim 7 - 8$ (see the next Section) may be significantly overestimated. At $z \sim 4 - 6$ however, our masses are not expected to change drastically. First, most of our massive sources at $z \sim 4 - 5$ would not be selected by Williams et al. (2024) because they have F150W < 27 mag, i.e. they are not faint enough in $F150W$ to be considered analogs of H-dropouts or HST-dark galaxies (see also Fig. 13). The biggest differences in masses inferred with and without MIRI-data are found for galaxies at $z \gtrsim 5$ (see fig. 2 in Williams et al. (2024)). Consistent with this, both Papovich et al. (2023) and Wang et al. (2024a) find only modest biases in the NIRCam-inferred stellar masses at $z \sim 4 - 6$.

Our inferred SMFs at $z \sim 4 - 6$ are therefore not only robust to LRD-contamination, but the masses and redshifts in this range also appear to be well-constrained and not subject to any significant biases based on the available NIRCam + HST data. The situation is however more complicated at higher redshifts $z \gtrsim 7$ which we discuss in the following.

### 4.4 Overly massive galaxies at $z \gtrsim 7$

As mentioned several times above, inferring stellar masses and redshifts from the photometry of extremely red galaxies is difficult and at the same time critical in order to constrain the high-mass end of the SMF which is dominated by (UV-)red galaxies. We have investigated the impact of LRDs with characteristic V-shaped SEDs as well as of red compact galaxies, i.e. galaxies that satisfy the red colour cut in the rest-optical as well as the compactness criterion for LRDs proposed in Labbe et al. (2023a), but are not or only marginally detected in the rest-UV. Those galaxies do not or only negligibly affect the SMFs at $z \sim 4 - 6$, but they matter more at $z \sim 7 - 8$ as can be seen in Figs 3, 4, and 6. To better understand the extreme objects that shape the high-mass end of the SMF, we have produced and visually inspected bagpipes SED-plots and *JWST* imaging cut-outs of all galaxies selected as LRDs, red compact, and of other galaxies that have $\log(M_*/M_\odot) > 10$. Among the red compact objects, there are 30 with $\log(M_*/M_\odot) > 10$. They can broadly be split into two categories: 16 of them show a purely red SED with no detections in the rest-frame UV. Those sources usually have a very broad P(z) and therefore poorly constrained masses which mitigates their effect on our SMFs due to the sampling of the posterior distributions. The other 14 red compact sources have marginal detections in the rest-frame UV, often only in one filter, hence they do not pass our LRD selection. However, the detections in the rest-UV are usually not reproduced by the red SED fitted by `bagpipes`. Together with their point-like morphology, this suggests that at least some of them are likely LRD AGNs. We show examples of each class of objects: an LRD, red compact, and massive galaxies in Figs B1 and B2 in Appendix B.

The visual inspection of all the massive galaxies ($M_* > 10^{10}\,\mathrm{M}_\odot$) confirmed that the vast majority of sources at $z \sim 4-6$ have plausible SED-fits and inferred masses. However, all the massive sources at $z \sim 7-9$ typically show a red continuum with poorly constrained redshift and mass, comparable to the source shown in the lower three panels of Fig. B1, and at least have a plausible lower redshift and thus lower mass solution.

The masses and redshifts of all galaxies with $M_* > 10^{10}\,\mathrm{M}_\odot$ at $z \sim 7\text{–}8$ may therefore be significantly overestimated. We emphasize again that our inferred number densities shown in this range are largely only upper limits, and that they are not included when deriving our Schechter fits.

### 4.5 The z ∼ 9 SMF

We also visually inspect the SEDs and *JWST* filter cut-outs of all the sources whose median redshift from `bagpipes` falls in our $z \sim 9$ bin. In total, there are 54 galaxies above our respective mass completeness limits in all fields combined. Of those, 25 have plausible fits and inferred masses; For 3 sources, their red SEDs, extreme inferred masses at $z \sim 9$, and/or detections below the suggested Lyman break indicate that they are likely to be at $z \sim 3$. The remaining 26 objects typically have poorly constrained redshifts and therefore stellar masses, showing an extremely broad and/or multiply peaked P(z). They are often red and/or relatively faint, and drop out of most of the shorter wavelength filters considered ($< 2\,\mu\mathrm{m}$). 10 of those objects have a very red optical continuum and a compact morphology, and 2 of them are classified as red compact (Section 2.5). In some cases, the inferred median redshift of $z \sim 9$ lies between two peaks in the P(z). Note however that due to our SMFs being *sampled* from the `bagpipes` posterior distributions, such sources will contribute to the inferred SMF data points in various bins, and therefore likely not have a big impact on the inferred number density in any specific bin. At $\log(M_*/\mathrm{M}_\odot) > 9.5$, we only find two plausible objects with well-constrained masses and redshifts. We conclude that our results for the SMF at $z \sim 9$ have to be interpreted cautiously, and are only reliable at $8.5 < \log(M_*/\mathrm{M}_\odot) < 9.5$. It is therefore possible that the SMF drops more rapidly towards higher masses than suggested by our best-fitting Schechter function which would alleviate the potential tension with cosmological constraints suggested in Fig. 11, bottom right panel, as well as the remarkably shallow evolution of the CSMD from $z \sim 8$ to $z \sim 9$ shown in Fig. 12.

## 5 SUMMARY AND CONCLUSIONS

We perform a detailed analysis of the stellar masses and UV-slopes, $\beta$, of galaxies at $z \sim 4-9$ based on public *JWST* + *HST* imaging data over the CEERS-EGS, PRIMER-UDS, PRIMER-COSMOS, and JADES GOODS-S survey fields. After providing a detailed description of the photometric catalogues that form the basis of this work, we select a sample of galaxies at $z \gtrsim 3$ and use `bagpipes` to infer their redshifts, stellar masses, and UV-slopes.

We match our sample to an LBG-sample selected in B15 over the same fields and perform a self-consistent comparison between an *HST*-based rest-frame UV detected, colour–colour selected sample of LBGs and a *JWST*-based rest-frame optically detected, photo-z selected sample of galaxies. Then, we split our sample into UV-red and UV-blue galaxies, adopting a simple cut in the UV-slope at $\beta = -1.2$, and investigate how the UV-red fraction evolves as a function of stellar mass and redshift. To assess the impact of (UV-)red galaxies that are only accessible through or at least better characterized with *JWST* compared to the pre-*JWST* era, we compute SMFs as well as the CSMD and compare to existing results in the (mostly pre-*JWST*) literature.

To provide further insights, we investigate the SMFs of UV-red and UV-blue galaxies as well as their contribution to the CSMD separately. This is followed by a detailed discussion of the contribution to the SMF of UV-red and supposed extremely massive galaxies in the context of recent findings with *JWST* regarding the so-called LRDs as well as in the context of existing research in the fields of sub-mm galaxies, *HST*-dark and OFGs.

Our findings can be summarized as follows:

(i) With *JWST*, we are detecting UV-red galaxies that were missing from *HST*-based LBG-samples. Those galaxies are typically massive ($M_* > 10^{10}\mathrm{M}_\odot$) and many of them have been detected with *Spitzer*/IRAC and/or at sub-mm wavelength prior to *JWST*. There is however a small number of elusive fainter UV-red sources, down to masses of $\log(M_*/\mathrm{M}_\odot) \sim 8$ which are exclusive to *JWST*.

(ii) The fraction of UV-red galaxies ($\beta > -1.2$) is a strong function of stellar mass at fixed redshift and shows no clear signs of evolving with redshift. Comparing to the UV-red fraction inferred from the subsample of galaxies present in B15 shows that there are no big differences in the mass range $\log(M_*/\mathrm{M}_\odot) \sim 8-10$ at $z \sim 4-6$. At $z \sim 7-8$, the B15 sample is however missing a significant fraction of UV-red galaxies at $\log(M_*/\mathrm{M}_\odot) \gtrsim 9$.

(iii) We find a remarkable number of 53 sources in our sample with stellar masses $\log(M_*/\mathrm{M}_\odot) > 10.5$ at $z \sim 4$. As many as $\sim 34$ per cent of those galaxies can be classified as '*HST*-dark' and would therefore be missing from typical *HST*-based LBG samples. The number of such massive sources becomes small at $z \gtrsim 5$. However, most of them are UV-red, and missing from the B15 sample.

(iv) Our inferred SMFs are broadly consistent within measurement uncertainties with the pre-*JWST* literature at all the redshifts probed. We do however identify four peculiarities of our SMFs: (1) a high galaxy number density at the high-mass end at $z \sim 4$, relative to literature results, (2) a modest steepening of the low-mass end slope of the SMFs from $z \sim 4$ to $z \sim 6$, reaching a low value of $\alpha \sim -2$ at $z \sim 6$, (3) an excess at the high-mass end at $z \sim 7-8$, driven by extremely red sources with highly uncertain masses and redshifts, and (4) a high abundance of galaxies at $z \sim 9$ compared to existing literature results.

(v) A more detailed comparison of the high-mass end at $z \sim 4$ to pre-*JWST* literature results shows that our inferred SMF is at the upper edge of existing predictions but still consistent with other studies, in particular if they derive the high-mass end of the SMF from *Spitzer*/IRAC or ground-based $K_S$-band-selected samples (e.g. Caputi et al. 2015; Weaver et al. 2023).

(vi) As a consequence, we find a strong evolution of the high-mass end of the SMF from $z \sim 4$ to $z \sim 5$, which is entirely driven by UV-red galaxies. The mass density of UV-red ($\beta > -1.2$) galaxies increases by a factor $\sim 8$ from $z \sim 5$ to $z \sim 4$ where UV-red galaxies start to dominate the total CSMD for $M_* > 10^8\,\mathrm{M}_\odot$, implying a rapid build-up of massive dusty systems in this redshift range.

(vii) The inferred excess of very massive galaxies at $z \sim 7-8$ is based on a small number of sources, and it is sensitive to the

inclusions of LRDs and red compact sources, i.e. objects that may be LRDs, but are simply too faint in the rest-UV to be detected there. While such sources significantly affect our sample at the massive end at $z \sim 7-8$, they have a negligible impact on our SMFs at lower redshifts.

(viii) At $z \sim 9$, we can only robustly constrain the SMF in the stellar mass range $8.5 < \log(M_*/M_\odot) < 9.5$ and with upper limits at higher masses, causing the corresponding Schechter fit to be uncertain. The measured densities nevertheless suggest a higher abundance of galaxies at $z \sim 9$ than previously measured, in particular with respect to Stefanon et al. (2021).

(ix) Comparing to theoretical SMFs, inferred from the HMF, assuming different values of $\epsilon$ suggests that an increasing efficiency towards higher redshifts is required to explain the high-mass end of the SMF, reaching values of $\epsilon \sim 0.3$ at $z \sim 7-8$.

(x) Our SMFs are remarkably consistent with SMFs from a wealth of simulations out to at least $z \sim 6$. At higher redshifts, the scatter between different simulations increases. The various reasons for, and implications of the observed differences remain to be investigated.

(xi) Integrating our SMFs to infer the CSMD yields results consistent with the literature, but at the upper edge of predictions, both at $z \sim 4$ and at $z \sim 8-9$. Our results suggest a rather steep evolution from $z \sim 4-6$ which then becomes much shallower at $z \sim 7-9$.

To make further progress beyond this work, larger NIRCam survey areas, complemented by some very deep pointings, possibly making use of gravitational lensing effects, will be needed to better constrain both the high-mass and the low-mass ends of the SMF at the redshifts probed here and beyond. Further, complementary data at longer wavelengths, which may be provided by MIRI, ALMA, or other facilities over some fraction of the NIRCam survey areas, will help to better constrain the stellar masses and understand the nature of ultramassive red sources observed at high redshifts.

## ACKNOWLEDGEMENTS

This work is based on observations made with the NASA/ESA/CSA *JWST*. The raw data were obtained from the Mikulski Archive for Space Telescopes at the Space Telescope Science Institute, which is operated by the Association of Universities for Research in Astronomy, Inc., under NASA contract NAS 5–03127 for *JWST*. The authors acknowledge the CEERS and PRIMER teams for developing their observing program with a zero-exclusive-access period. This research is based on observations made with the NASA/ESA *Hubble Space Telescope* obtained from the Space Telescope Science Institute (STScI), which is operated by the Association of Universities for Research in Astronomy, Inc., under NASA contract NAS 5–26555.

The imaging data products presented herein were retrieved from the Dawn *JWST* Archive (DJA). DJA is an initiative of the Cosmic Dawn Center, which is funded by the Danish National Research Foundation (DNRF140).

This work has received funding from the Swiss State Secretariat for Education, Research and Innovation (SERI) under contract number MB22.00072, as well as from the Swiss National Science Foundation (SNSF) through project grant 200020_207349.

RG gratefully acknowledges support from the Inlaks Shivdasani Foundation. JSD acknowledges the support of the Royal Society through the award of a Royal Society Research Professorship. DM acknowledges funding from *JWST*-GO-01895.013, provided through a grant from the Space Telescope Science Institute under NASA contract NAS5-03127. RB acknowledges support from an Science and Technology Facilities Council (STFC) Ernest Rutherford Fellowship (grant number ST/T003596/1). PD acknowledges support from the Dutch Research Council (NWO) grant 016.VIDI.189.162 (ODIN) and warmly thanks the European Commission's and University of Groningen's CO-FUND Rosalind Franklin program. MS acknowledges support from the European Research Commission Consolidator Grant 101088789 (SFEER), from the CIDEGENT/2021/059 grant by Generalitat Valenciana, and from project PID2019-109592GB-I00/AEI/10.13039/501100011033 by the Spanish Ministerio de Ciencia e Innovación – Agencia Estatal de Investigación. FC acknowledges support from a UK Research and Innovation (UKRI) Frontier Research Guarantee Grant (PI F. Cullen; grant reference: EP/X021025/1). RPN acknowledges support for this work provided by NASA through the NASA Hubble Fellowship grant HST-HF2-51515.001-A and JWST program #2279 awarded by the Space Telescope Science Institute, which is operated by the Association of Universities for Research in Astronomy, Incorporated, under NASA contract NAS5-26555.

## DATA AVAILABILITY

The high-level science product images used here are v7 reductions publicly available from the DAWN *JWST* archive (DJA) at https://dawn-cph.github.io/dja/imaging/v7/. Photometric catalogues of all the sources and physical properties, as well as machine-readable versions of Tables 2–4 are available at https://zenodo.org/records/13166668.

## APPENDIX A: FIELD BY FIELD SMFS

To provide an overview over the field to field variations of our SMFs, we present the SMFs inferred from each of the four fields used in this work individually in Fig. A1. They are computed in analogy to the SMFs shown in e.g. Fig. 6, and excluding LRDs according to Section 2.5. We also derive Schechter function fits in each field. For these fits, we fix $M^*$ to the value inferred (or fixed) for the full SMF as described in Section 3.2.2 in all redshift bins, since the high-mass end of the SMF is poorly constrained by any individual field. The scatter among the different SMFs gives a rough idea of the cosmic variance involved in these measurements. However, the scatter among individual fields is expected to be larger than the cosmic variance affecting the SMF inferred from all four fields combined. Further, an empirical estimate of cosmic variance is complicated by the different survey areas, depths, and survey geometries of the four fields, which is why we do not attempt to measure it here. We note however that the simple scatter among the SMF-values inferred for individual fields is comparable to or even smaller than the applied cosmic variance uncertainties based on Moster et al. (2011), which therefore seems to give rather large values.

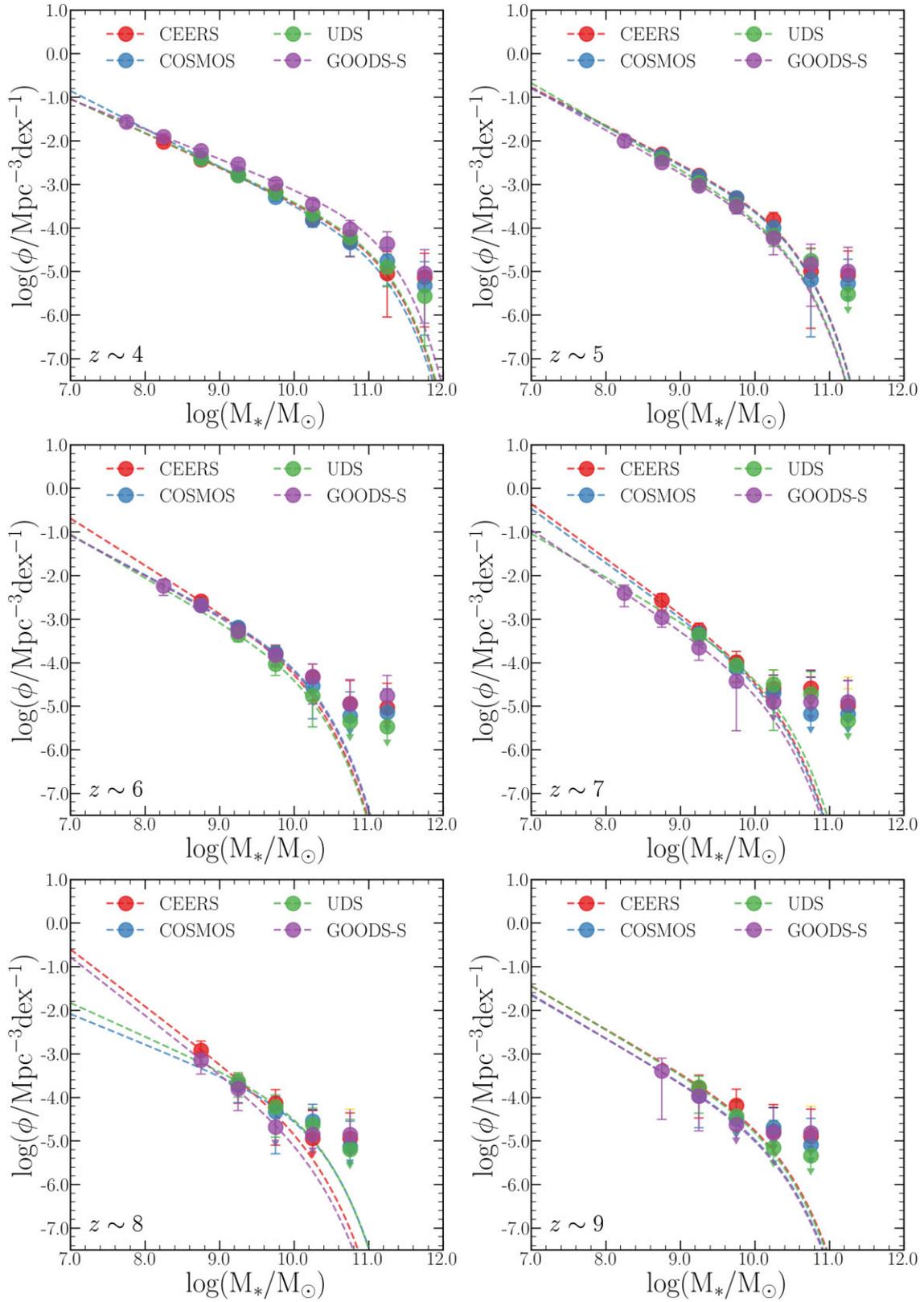

**Figure A1.** SMFs for each of the four fields (CEERS, PRIMER-COSMOS, PRIMER-UDS, and GOODS-S) separately in six different redshift bins. The dashed lines represent Schechter function fits to the SMFs in each field and redshift bin. In these fits, the parameter $M^*$ is fixed to the value inferred or set for the full SMF according to Table 4.

# APPENDIX B: EXAMPLE SEDS OF MASSIVE GALAXIES

In Fig. B1, we show a typical bright LRD in the top panel with clear detections in the rest-frame UV that are not matched by the best-fitting SED from `bagpipes` and an unreasonably high inferred stellar mass of almost $10^{11}$ M$_\odot$ at $z \sim 8$. The second panel shows a source selected as red compact which only has a marginal detection in $F150W$ in the rest-UV and is therefore not selected as an LRD. The two sources in the bottom panels show an extremely red SED and drop out of all filters at $\lesssim 2\mu$m, leaving any possible rest-UV flux unconstrained and yielding an implausibly massive solution at $z \sim 7$ (third panel) or an almost completely unconstrained P(z) (and therefore stellar mass) over the range $z \sim 3 - 10$ (bottom panel). The displayed sources with ID 42368 (PRIMER-COSMOS) and ID 3133 (PRIMER-UDS) have already been published in Ashby et al. (2015).

Another example of a supposedly extremely massive galaxy at $z \sim 8$ is shown in the top panel of Fig. B2. In this case, we argue the secondary solution at $z \sim 3$ to be much more plausible. In the middle two panels in Fig. B2, we show two massive galaxies at $z \sim 4$, representing the two types of galaxies contributing to the high-mass end in this redshift bin. ID 102915 in PRIMER-COSMOS is a beautiful face-on dusty spiral galaxy with a redshift of $z \sim 3.54$ and a stellar mass close to $10^{11}$ M$_\odot$, and ID 118183 in PRIMER-UDS is a quiescent galaxy with a similar mass. Finally, ID 133948 in PRIMER-UDS (bottom panel in Fig. B2) is plausibly fit at $z \sim 7.57$ with a remarkable $M_* = 10^{10.42}$ M$_\odot$. The photometry is consistent with a slightly lower redshift of $z \sim 7$, which would cause the mass of this source to be somewhat less extreme but still $> 10^{10}$ M$_\odot$. All the sources shown in Fig. B2 have already been published in Ashby et al. (2013).

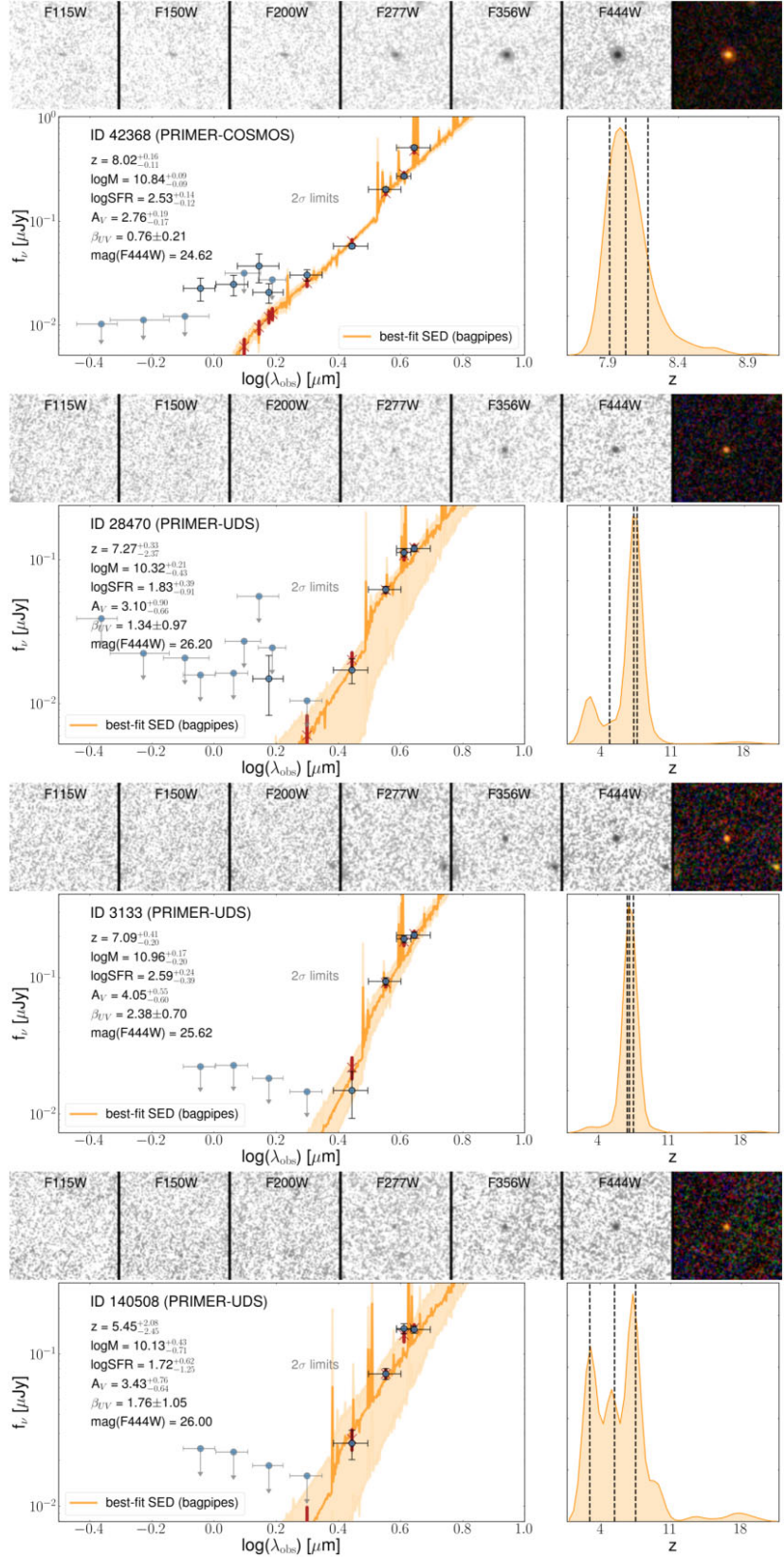

**Figure B1.** Example SEDs of galaxies selected as LRDs (top panel) and as red compact sources (remaining three panels). The source in the second panel shows a marginal detection in the rest-UV (F150W) and might well be an LRD-type source. The sources in the two bottom panels show an extremely red continuum and drop out of all filters at $\lesssim 2\,\mu$m, yielding implausibly massive solutions at high redshift that cannot be ruled out based on the available data (third panel) or almost completely unconstrained masses and redshifts (bottom panel).

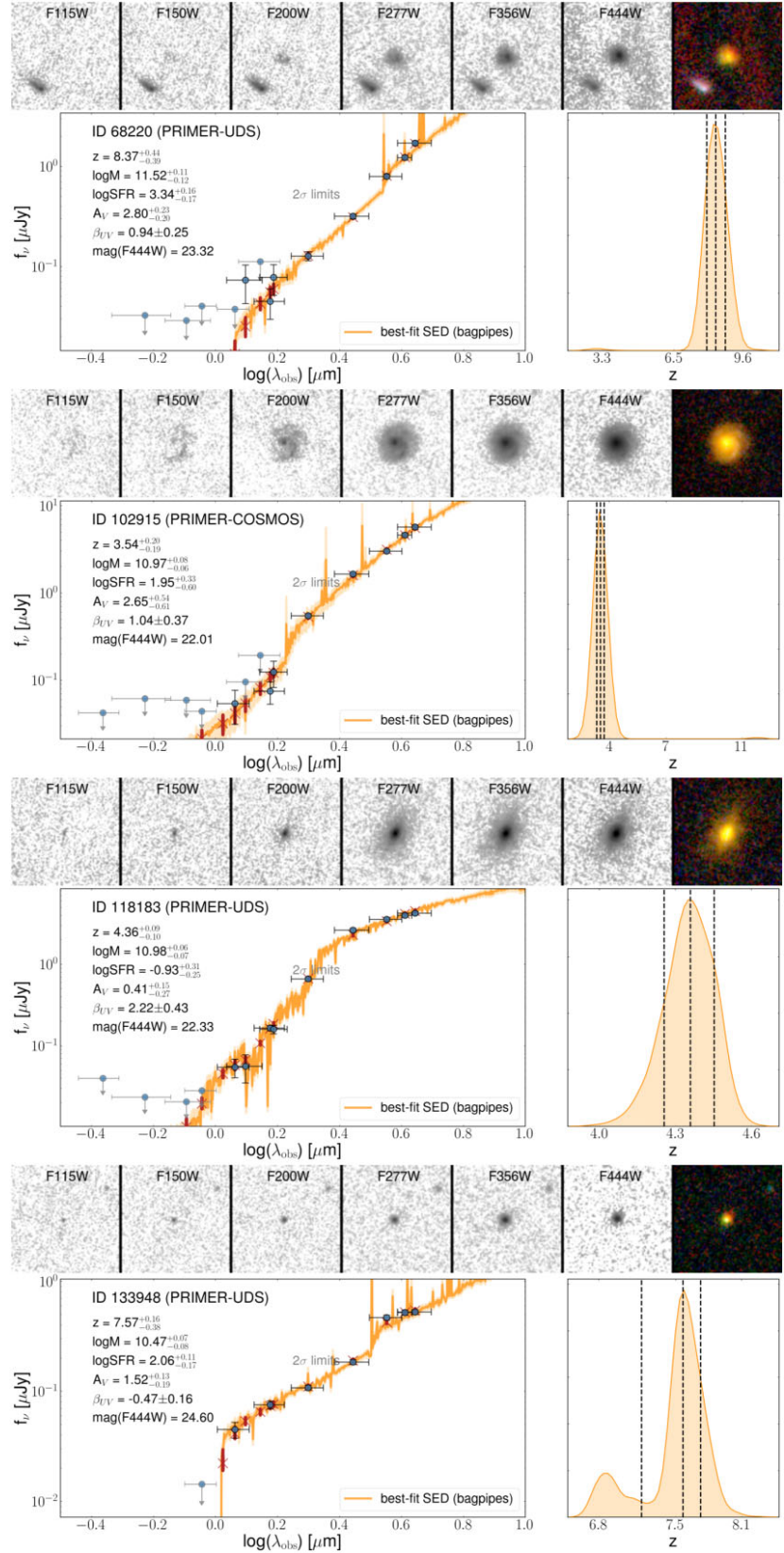

**Figure B2.** Example SEDs of massive galaxies. In the top panel, we show an example of a suggested extremely massive galaxy at $z \sim 8$, where we argue the secondary $z \sim 3$ solution to be more plausible. The second panel shows a massive dusty spiral galaxy at $z \sim 3.5$, and the third panel a massive quiescent galaxy at $z \sim 4.4$. Finally, the bottom panel shows our most plausible very massive galaxy in the range $z \sim 7-8$.

[1]*Department of Astronomy, University of Geneva, Chemin Pegasi 51, CH-1290 Versoix, Switzerland*
[2]*Cosmic Dawn Center (DAWN)*
[3]*Niels Bohr Institute, University of Copenhagen, Jagtvej 128, DK-2200 Copenhagen N, Denmark*
[4]*Institute for Astronomy, University of Edinburgh, Royal Observatory, Edinburgh EH9 3HJ, UK*
[5]*Department of Physics and Astronomy, University College London, Gower St, London WC1E 6BT, UK*
[6]*INAF - Osservatorio Astronomico di Roma, via di Frascati 33, I-00078 Monte Porzio Catone, Italy*
[7]*Department of Astronomy, University of Massachusetts Amherst, Amherst, MA 01003, USA*
[8]*Leiden Observatory, Leiden University, NL-2300 RA Leiden, the Netherlands*
[9]*Jodrell Bank Centre for Astrophysics, University of Manchester, Oxford Road, Manchester, M13 9PL, UK*
[10]*Kapteyn Astronomical Institute, University of Groningen, NL-9700 AV Groningen, the Netherlands*
[11]*NSF's National Optical-Infrared Astronomy Research Laboratory (NOIRLab), 950 N. Cherry Ave., Tucson, AZ 85719, USA*
[12]*Space Telescope Science Institute, 3700 San Martin Drive, Baltimore, MD 21218, USA*
[13]*Department of Astronomy and Astrophysics, University of California, Santa Cruz, CA 95064, USA*
[14]*Centre for Astrophysics and Supercomputing, Swinburne University of Technology, Melbourne, VIC 3122, Australia*
[15]*Department of Physics and Astronomy, Tufts University, Medford, MA 02155, USA*
[16]*MIT Kavli Institute for Astrophysics and Space Research, 77 Massachusetts Avenue, Cambridge, MA 02139, USA*
[17]*Centro de Astrobiología (CAB), CSIC-INTA, Ctra. de Ajalvir km 4, Torrejón de Ardoz, E-28850 Madrid, Spain*
[18]*Departament d'Astronomia i Astrofísica, Universitat de València, C. Dr Moliner 50, E-46100 Burjassot València, Spain*
[19]*Unidad Asociada CSIC 'Grupo de Astrofísica Extragaláctica y Cosmología' (Instituto de Física de Cantabria - Universitat de València)*

This paper has been typeset from a T<sub>E</sub>X/LAT<sub>E</sub>X file prepared by the author.